\documentclass{aa}  

\usepackage{graphicx}
\usepackage{newtxtext,newtxmath}
\usepackage[T1]{fontenc}

\DeclareRobustCommand{\VAN}[3]{#2}
\let\VANthebibliography\thebibliography
\def\thebibliography{\DeclareRobustCommand{\VAN}[3]{##3}\VANthebibliography}

\usepackage{graphicx}
\usepackage{amsmath}
\usepackage{hyperref}
\usepackage{tablefootnote}
\usepackage{siunitx}
\usepackage{threeparttable}
\hypersetup{colorlinks,linkcolor={blue},citecolor={blue},urlcolor={blue}} 
\usepackage{orcidlink}

\usepackage[tableposition=top]{caption}
\usepackage[protrusion=true,expansion]{microtype}
\usepackage{color}

\newcommand{\oiii}{[O\,\textsc{iii}]}
\newcommand{\nii}{[N\,\textsc{ii}]}
\newcommand{\sii}{[S\,\textsc{ii}]}

\newcommand{\siii}{[S\,\textsc{iii}]}
\newcommand{\oii}{[O\,\textsc{ii}]}

\newcommand{\hii}{H\,\textsc{ii}}

\begin{document} 

   \title{SDSS-V LVM: Resolving Physical Conditions in the Trifid Nebula}

    % might not be the final order
    
   \author{
    Natascha Sattler\inst{1}\thanks{\email{nataschasattler@outlook.de}}\orcidlink{0000-0002-8883-6018}
    \and
    J. Eduardo M\'endez-Delgado\inst{1,2}\orcidlink{0000-0002-6972-6411}
    \and
    Kathryn Kreckel\inst{1}
    \and
    Christophe Morisset\inst{3, 4}\orcidlink{0000-0001-5801-6724}
    \and
    Oleg Egorov\inst{1}
    \and
    Evgeniya Egorova\inst{1}
    \and
    Ahmad Nemer\inst{5,6}
    \and
    Fu-Heng Liang\inst{1}\orcidlink{0000-0003-2496-1247}
    \and
    A.A.C. Sander\inst{1,7}\orcidlink{0000-0002-2090-9751}
    \and
    Alexandre Roman-Lopes\inst{8}\orcidlink{0000-0002-1379-4204}
    \and
    Carlos G. Rom\'an-Z\'u\~niga\inst{3}\orcidlink{0000-0001-8600-4798}
    \and
    Evelyn J. Johnston\inst{9}\orcidlink{0000-0002-2368-6469}
    \and
    Sebasti\'an F.\ S\'anchez\inst{3,10}
    \and
    Jos\'e G. Fern\'andez-Trincado\inst{11}
    \and
    Niv Drory\inst{12}\orcidlink{0000-0002-7339-3170}
    \and
    Amrita Singh\inst{13}
    \and
    Dmitry Bizyaev\inst{14}
    \and
    Sumit K. Sarbadhicary\inst{15, 16, 17}\orcidlink{0000-0002-6313-4597}
    \and
    Pablo Garc\'ia\inst{18, 19}\orcidlink{0000-0002-8586-6721}
    \and
    Alfredo Mej\'{\i}a-Narv\'aez\inst{20}\orcidlink{0000-0002-8931-2398}
    \and
    Guillermo A. Blanc\inst{13,21}\orcidlink{0000-0003-4218-3944}
    }
   \institute{
    %1
    Astronomisches Rechen-Institut, Zentrum f\"{u}r Astronomie der Universit\"{a}t Heidelberg, M\"{o}nchhofstra\ss e 12-14, D-69120 Heidelberg, Germany
    \and
    %2
    Instituto de Astronom\'{\i}a, Universidad Nacional Aut\'onoma de M\'exico, Ap. 70-264, 04510 CDMX, Mexico
    \and
    %3
    Instituto de Astronom\'ia, Universidad Nacional Aut\'onoma de M\'exico, Ap. 106, 22800 Ensenada, BC, M\'exico
    \and
    %4
    Instituto de Ciencias F\'{\i}sicas, Universidad Nacional Aut\'onoma de M\'exico, Av. Universidad s/n, 62210 Cuernavaca, Morelos, Mexico
    \and
    %5
    New York University Abu Dhabi, PO Box 129188, Abu Dhabi, UAE
    \and
    %6
    Center for Astrophysics and Space Science, NYU Abu Dhabi, PO Box 129188, Abu Dhabi, UAE
    \and
    %7
    Universit\"at Heidelberg, Interdiszipli\"ares Zentrum f\"ur Wissenschaftliches Rechnen, 69120 Heidelberg, Germany
    \and
    %8
    Department of Astronomy / Departamento de Astronomía, Universidad de La Serena, La Serena, Chile
    \and
    %9
    Instituto de Estudios Astrof\'isicos, Facultad de Ingenier\'ia y Ciencias, Universidad Diego Portales, Av. Ej\'ercito Libertador 441, Santiago, Chile
    \and
    %10
    Instituto de Astrof\'\i sica de Canarias, La Laguna, Tenerife, E-38200, Spain
    \and
    %11
    Universidad Cat\'olica del Norte, N\'ucleo UCN en Arqueolog\'ia Gal\'actica - Inst. de Astronom\'ia, Av. Angamos 0610, Antofagasta, Chile
    \and
    %12
    McDonald Observatory, The University of Texas at Austin, 1 University Station, Austin, TX 78712-0259, USA
    \and
    %13
    Departamento de Astronom\'{i}a, Universidad de Chile, Camino del Observatorio 1515, Las Condes, Santiago, Chile
    \and
    %14
    Apache Point Observatory and New Mexico State University, P.O. Box 59, Sunspot, NM, 88349-0059, USA
    \and
    %15
    Department of Physics and Astronomy, The Johns Hopkins University, Baltimore, MD 21218, USA
    \and
    %16
    Department of Physics, The Ohio State University, Columbus, Ohio 43210, USA
    \and
    %17
    Center for Cosmology \& Astro-Particle Physics, The Ohio State University, Columbus, Ohio 43210, USA
    \and
    %18
    Chinese Academy of Sciences South America Center for Astronomy, National Astronomical Observatories, CAS, Beijing 100101, China
    \and
    %19
    Instituto de Astronom\'ia, Universidad Cat\'olica del Norte, Av. Angamos 0610, Antofagasta, Chile
    \and
    %20
    Universidad de Chile, Av. Libertador Bernardo O'Higgins 1058, Santiago de Chile
    \and
    %21
    Observatories of the Carnegie Institution for Science, 813 Santa Barbara Street, Pasadena, CA 91101, USA
    }

  \abstract
  {}{The chemical abundance of the interstellar medium sets the initial conditions for star formation and provides a probe of chemical galaxy evolution models. 
  However, unresolved inhomogeneities in the electron temperature can lead to a systematic underestimation of the abundances. 
  We aim to directly test this effect.}
  {We use the SDSS-V Local Volume Mapper to spatially map the physical conditions of the Trifid Nebula (M\,20), a Galactic \hii\ region ionized by a single mid-type O star, at 0.24 pc resolution. 
  We exploit various emission lines (e.g., Hydrogen recombination lines and collisionally excited lines, including also faint auroral lines) and compute spatially resolved maps of \oii\ and \sii\ electron densities; \nii, \oii, \sii, \siii\ electron temperatures; and the ionic oxygen abundances.} 
  {We find internal variations of electron density that result from the ionization front, along with a negative radial gradient. 
  However, we do not find strong gradients or structures in the electron temperature and the total oxygen abundance, making the Trifid Nebula a relatively homogeneous \hii\ region at the observed spatial scale. 
  We compare these spatially resolved properties with equivalent integrated measurements of the Trifid Nebula and find no significant variations between integrated and spatially resolved conditions.} 
  {This isolated \hii\ region, ionized by a single O-star, represents a test case of an ideal Strömgren sphere. 
  The physical conditions in the Trifid Nebula behave as expected, with no significant differences between integrated and resolved measurements.} 

   \keywords{ISM: abundances -- ISM: general -- ISM: HII regions -- ISM: structure -- ISM: individual objects: M\,20 -- Galaxy: local interstellar matter}

   \maketitle

%%%%%%%%%%%%%%%%%%%%%%%%%%%%%%%%%%%%%%%%%%%%%%%%%%%%%%%%%%%%%%%%%%%%%%%%%%%%%%%%%%%%%%%%%%%%%%%%%%%%%%%%%%%%%%%%%%%%%%%

\section{Introduction}
\label{sec:introduction}

% chemical evolution
During the process of galaxy evolution, gas from the interstellar medium (ISM) is enriched by stellar feedback from the current generation of stars, through supernovae explosions and stellar winds \citep[e.g.,][]{gallart_2019, martig_2021, conroy_2022}. 
This enriched gas mixes into the ISM of the galaxy and eventually forms a new generation of stars with higher metal abundances \citep[e.g.,][]{worthey_1992, mo_2010, peletier_2013}.
Therefore, to understand galaxy formation and test models of chemical galaxy evolution, we need to determine the current elemental abundances, as they contain a record of the galaxy's formation history \citep[e.g.,][]{Peimbert2017, Maiolino2019}.
% H II region
This can be achieved by measuring the current chemical gas-phase abundances of star-forming \hii\ regions via bright optical emission lines \citep[e.g.,][]{Perez-Montero2017, Peimbert2017, Kewley2019a}.
\hii\ regions are primarily composed of hydrogen ionized by young massive stars (e.g., O-type stars) with a high rate of hydrogen-ionizing photon emission.
In addition, these regions also contain helium and heavier elements such as oxygen, nitrogen, and sulfur \citep[e.g.,][]{Esteban2017, Esteban2025}.

% heating and cooling
The temperature of the gas in the \hii\ region is governed by a balance of heating and cooling processes. 
The ISM gas is mainly heated by photoionization from the stellar ionizing sources, and cooled through recombination, free-free radiation, and collisionally excited line (CEL) radiation \citep{Osterbrock2006, Peimbert2017}. 
The free-free cooling is dominated by the hydrogen ions, due to their abundance in the ISM.
On the other hand, heavier ions such as oxygen or nitrogen contribute significantly to the cooling via CEL radiation because they have low-lying energy levels that are similar in energy to the kinetic energy of free electrons. 
This allows these heavier ions to be easily excited through collisions, unlike hydrogen and helium, whose excitation potentials are much higher.
Therefore, each element contributes to a different cooling mechanism, which makes the equilibrium temperature of the gas in the \hii\ region strongly dependent on the abundances of the different elements.

% emission lines
During these cooling processes, various kinds of emission lines are produced in the nebula.
The recombination cooling is responsible for the observed recombination lines (RL) that are primarily produced by hydrogen, giving rise to the Balmer, Paschen, and other hydrogen emission series.
Cooling through collisionally excited lines (CEL) in the optical mostly occurs via transitions that are forbidden in the electric dipole approximation but can proceed through less probable magnetic dipole (M1) or electric quadrupole (E2) mechanisms. 
Well-known examples include the strong \oiii$\lambda$5007 (M1) line and the weaker \oiii$\lambda$4931 (E2) line. 
The very low densities in ionized nebulae prevent collisional de-excitation, allowing these forbidden lines to become efficient cooling channels \citep{Baker1938}.
The relative fluxes of different emission lines are sensitive to the electron density, electron temperature, and the ionic abundances, hence can be used as a diagnostic of the \hii\ region's physical conditions \citep{Osterbrock2006, Draine2011}.

% inhomogeneities
It is often assumed that \hii\ regions are homogeneous systems \citep[e.g.,][]{Filippenko1985, Osterbrock1989}; however, the density and temperature distributions across observed \hii\ regions are not uniform, and spatial variations or inhomogeneities are measured \citep[e.g.,][]{Copetti2000, Malmann2002}.
% density structure
Most \hii\ regions show filamentary or shell-like electron density structures \citep[e.g.,][]{Kennicutt1984a}, and show a variety of radial gradients in density \citep[e.g.,][]{Binette2002, Herrera-Camus2016, Rubin2016}.
% temperature variations
Similarly, the electron temperature also shows inhomogeneities that have been theorized to cause underestimations of the ionic abundances \citep[e.g.,][]{Peimbert2004, Hagele2006, Peimbert2017, Kewley2019b}. 
However, measuring the electron temperature and its variations is challenging, as the measurements rely on faint auroral lines that are difficult to detect \citep[e.g.,][]{Hagele2006, Kewley2019a}.
These variations may come from shock waves and turbulence \citep{Peimbert1991, O'Dell2015, Arthur2016, Royer2025}, stellar winds from planetary nebula \citep{Peimbert1995a} or Wolf-Rayet stars \citep{Gonzalez-Delgado1994}, inhomogeneities in density or the distribution of multiple ionization sources \citep{Arthur2006, O'Dell2017}.
Controversially, some studies find no evidence of temperature fluctuations in single nebula \citep[e.g.,][]{Liu2006a, Stasinska2013b}.

% abundances
Internal variations of electron density and temperature may affect the calculation of the intrinsic ionic abundances, as temperature inhomogeneities are known to lead to an underestimation in the calculation of abundances, if not taken into account properly \citep{Peimbert2017, Mendez-Delgado2022b, Mendez-Delgado2023a, Mendez-Delgado2023b}.
% need spatially resolved data
Particularly, in integrated extra-galactic \hii\ regions or long-slit observations, the temperature inhomogeneities cannot be resolved and need to be estimated by comparing different ionic temperature measurements. 
This is because ions with different ionization potentials dominate in distinct regions of the nebula, so their temperatures probe different physical zones \citep{Peimbert2017}.
Since abundance determinations depend on these ionic temperatures, unresolved observations can introduce systematic biases in the derived ionic abundances. 
Therefore, we need spatially resolved observations of \hii\ regions to directly measure the extent of density and temperature inhomogeneities and obtain more accurate abundances.

% IFUs in the past
Previous approaches using integral field unit (IFU) data to study the detailed structures of \hii\ regions were done by e.g., \citet{Sanchez2007, Garcia-Benito2010, Relano2010, Lopez-Hernandez2013, Kumari2017, Dopita2019, Jin2023, Garner2025, Royer2025}.
For four \hii\ regions in the Large and Small Magellanic Cloud at a sampling of around 0.24 to 0.3\,pc, \citet{Jin2023} found negative radial gradients in electron density as well as positive and negative radial gradients in electron temperature.
However, the gradients in electron temperature are smaller than the uncertainties, so the measured temperatures across the four nebulae are consistent with being flat.
Moreover, \citet{Royer2025} find complex patterns of electronic density and temperature variations across the \hii\ region Sh2-158.

Using the Sloan Digital Sky Survey-V (SDSS-V) Local Volume Mapper (LVM), which is a recently commissioned spectroscopic survey \citep{Perruchot2018, Kollmeier2019, Drory2024, Blanc2024, Herbst2024, Kollmeier2025}, a homogeneous spatial sampling of multiple \hii\ regions and the ISM across the local galactic environment with high-quality spectra can be achieved.
The recent study of \citet{Kreckel2024} demonstrated the great potential of this survey, as it presented a first analysis of a larger region in the Orion nebula.\\

\begin{figure}
    \centering
    \includegraphics[width=1\columnwidth]{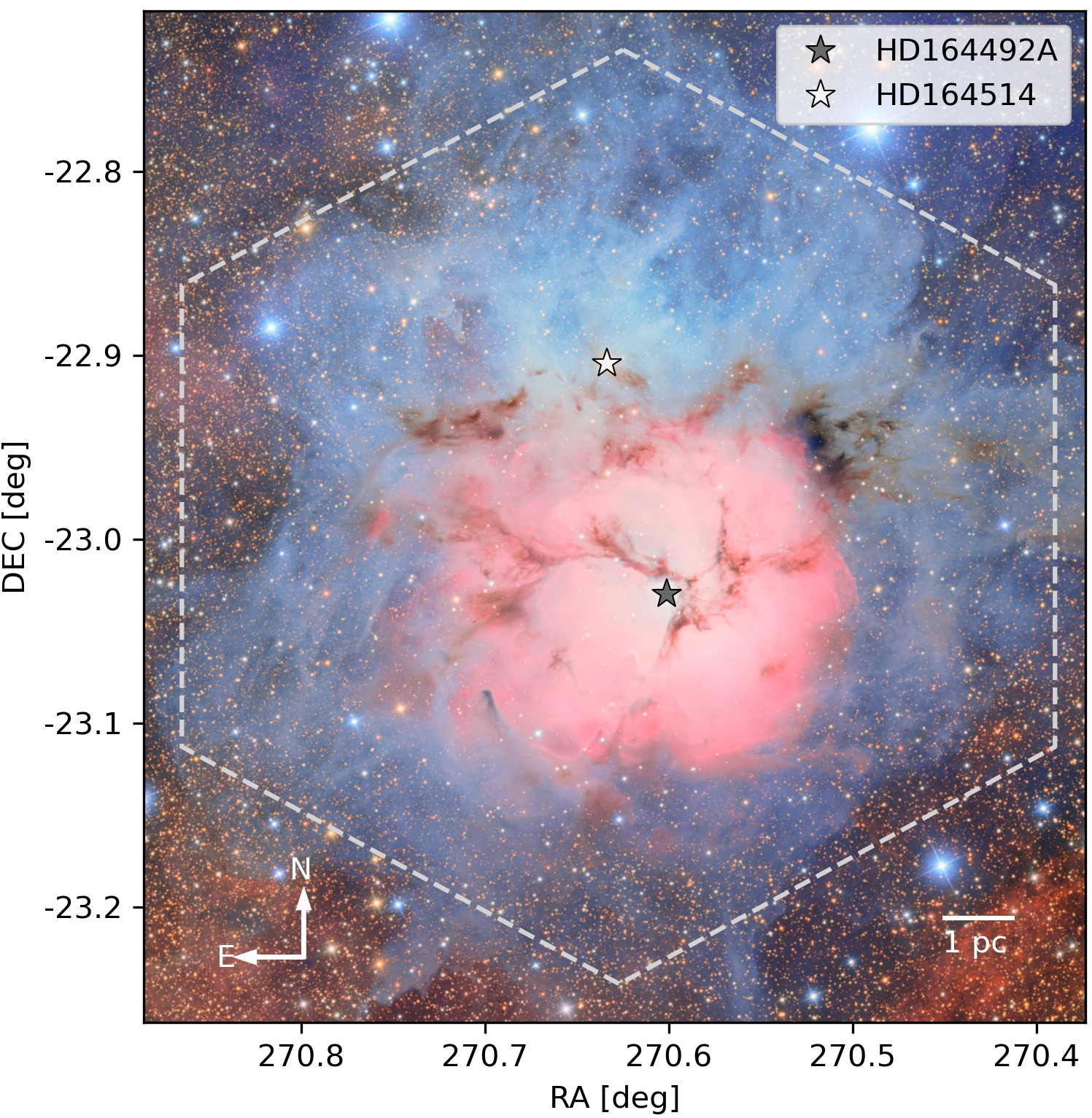}
    \caption{NSF–DOE Vera C. Rubin Observatory image of the Trifid Nebula, using LSST's (Legacy Survey of Space and Time, \citealt{Ivezic2019}) six filters: \textit{u}, \textit{g}, \textit{r}, \textit{i}, \textit{z} and \textit{y}.
    It shows the glowing pink emission nebula and the cool blue reflection nebula. 
    The position of the central ionizing source HD\,164492A is marked with a grey star. 
    The center of Trifid's reflection nebula \citep{Lynds1986} is marked by the A\,7 super-giant HD\,164514 (white star).
    The extent of the hexagonal LVM pointing is indicated by a grey dashed line.}
    %Digitized Sky Survey (DSS) 
    %Other massive candidate OB stars located in the Trifid region, identified by \citet{Povich2017}, are marked with blue diamonds.}
    \label{fig:rubin}
\end{figure}

% M20
The Trifid Nebula (M\,20), shown in Fig.~\ref{fig:rubin} from an NSF–DOE Vera C. Rubin Observatory image,\footnote{\hyperlink{https://rubinobservatory.org/news/first-imagery-rubin}{https://rubinobservatory.org/news/first-imagery-rubin}} is a small and nearly symmetrical \hii\ region ionized by HD\,164492A, classified as an O\,7.5~V star \citep{Sota2014}.
However, its measured effective temperature of 38000~K \citep{Martins2015} suggests a slightly earlier type, around O\,6.5\,V.
With the ionization dominated by this single O-type star, M\,20 offers an ideal laboratory for studying spherical symmetry within an individual Strömgren sphere \citep{Stromgren1939}.

% stellar properties
HD\,164492A exhibits a soft and variable X-ray emission spectrum with $kT \approx 0.5$~keV and $\log L_\text{X}/L_\text{bol} < -7$, likely produced by wind-intrinsic shocks \citep{Rho2004}.
This system was investigated for signs of binarity by \citet{Stickland2001}, but they could not find conclusive insights. 
While we cannot rule out that HD\,164492A is a binary, the available IUE UV spectrum does not show obvious signs of a significantly hotter or cooler companion. 
While a more thorough investigation would be needed to set stronger constraints on multiplicity, we do not expect that this would lead to a drastically different result in the ionizing photon budget.
In addition to the central ionizing star, younger lower-mass stellar populations (<~1~Myr) are present in M\,20 \citep{Tapia2018} as well as several protostars \citep{Rho2001, Rho2006, Yusef-Zadeh2005a}, indicating ongoing star formation.

% distance
In this study, we adopt a distance to M\,20 of 1.42~$^{+0.09}_{-0.08}$~kpc calculated by \citet{Mendez-Delgado2022a} using Gaia EDR3 parallaxes \citep{GaiaCollaboration2020} of the central stellar system HD\,164492, containing HD\,164492A.
% reflection nebula
A reflection nebula, caused by scattered light from the A\,7 super-giant HD\,164514 \citep{Lynds1986} with a Gaia EDR3 distance of 1.191~$\pm$~0.047~kpc \citep{GaiaCollaboration2020}, is located north of the Trifid Nebula.
\citet{Lynds1986} suggested that the grains in the reflection nebula have a higher albedo than the particles in the Trifid Nebula itself, implying that the O-star HD\,164492A must have modified the nature of the surrounding grains.

% formation
The current generation of stars in M\,20 likely formed about 1~Myr ago following a collision between two molecular clouds \citep{Torii2011, Torii2017, Fukui2021, Kalari2021}, where the ionizing O-type star HD\,164492A presumably formed around 0.3~Myr ago \citep{Torii2017}.
However, other studies by \citet{Fukuda2000, Rho2008, Kuhn2022} suggested that the kinematic features of the clouds may not necessarily come from a cloud-cloud collision but can also be interpreted as the expansion of the \hii\ region or turbulence.\\

% this paper
In this paper, we use integral field spectroscopic data from the LVM to spatially map the physical conditions of the Trifid Nebula at 0.24\,pc resolution.
We organize this study as follows:
First, we describe the observations as well as the data reduction in Sect.~\ref{sec:observations}.
Next, we describe the analysis methods to infer physical properties in Sect.~\ref{sec:analysis}.
Then, we present our results of the electron density, temperature, and ionic abundances in Sect.~\ref{sec:results}.
Further, the results are discussed in Sect.~\ref{sec:discussion} and our conclusions are given in Sect.~\ref{sec:conclusion}.

\begin{figure*}
    \centering
    \begin{minipage}[t]{0.95\textwidth}
        \centering
        \includegraphics[width=\textwidth]{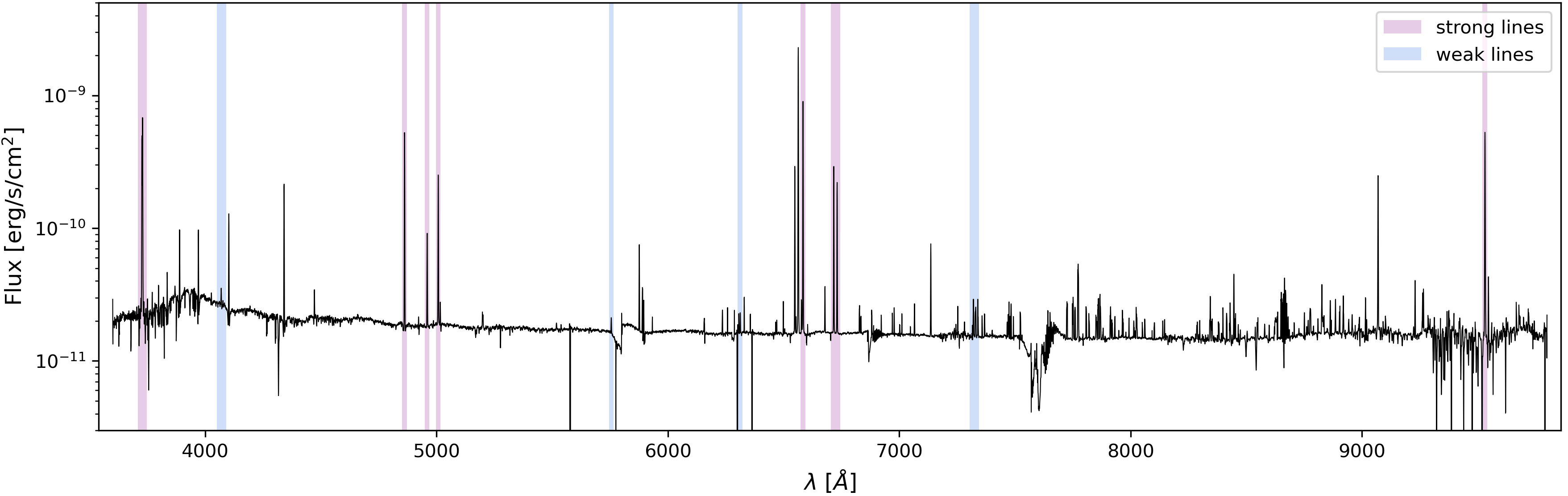}
    \end{minipage}
    \begin{minipage}[b]{\textwidth}
        \begin{minipage}[l]{0.5\textwidth}
            \hspace{0.71cm}
            \includegraphics[width=1.12\textwidth]{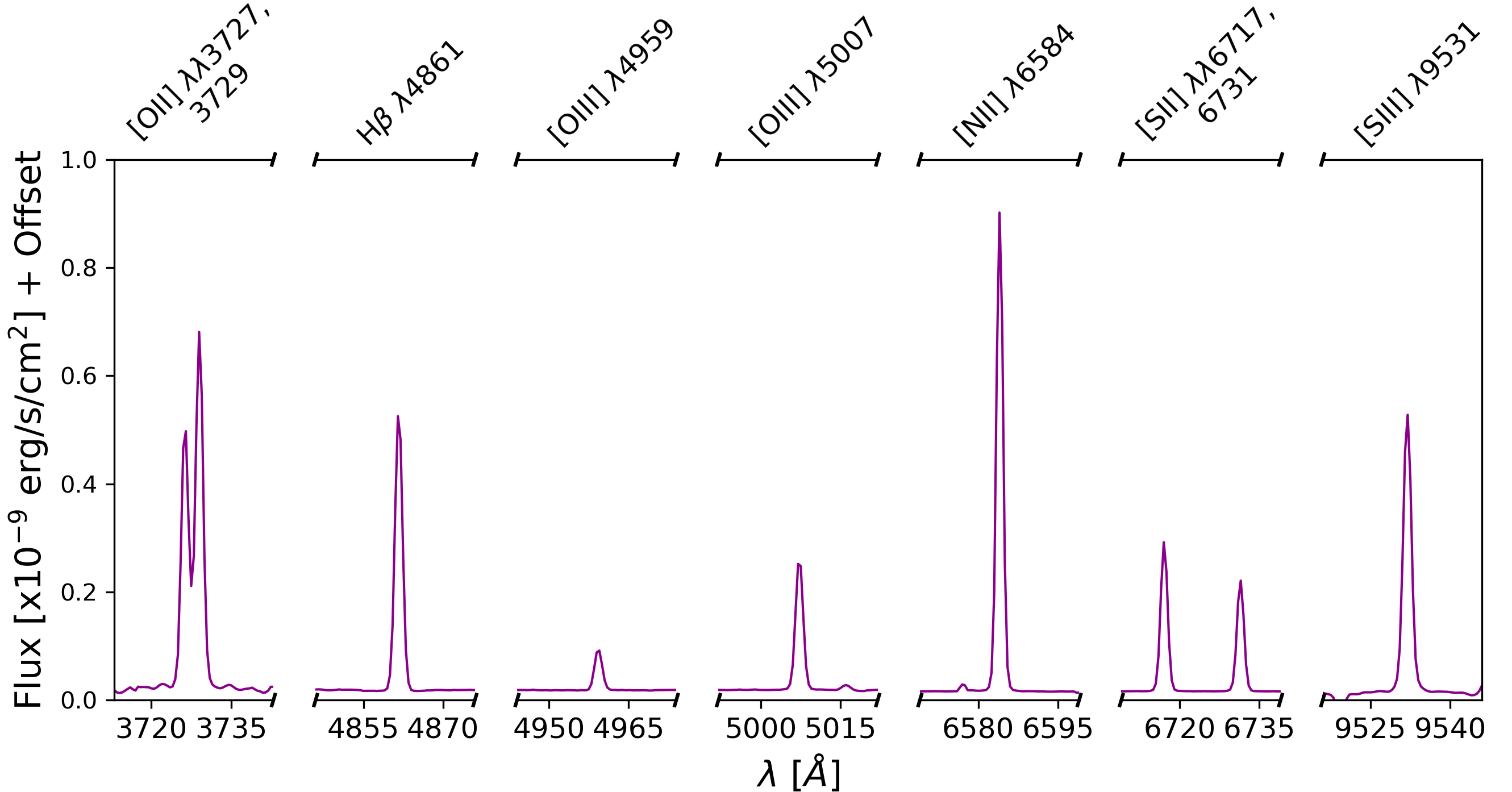}
        \end{minipage}
        \hspace{-0.2cm}
        \begin{minipage}[r]{0.5\textwidth}
            \hspace{2.8cm}
            \includegraphics[width=0.63\textwidth]{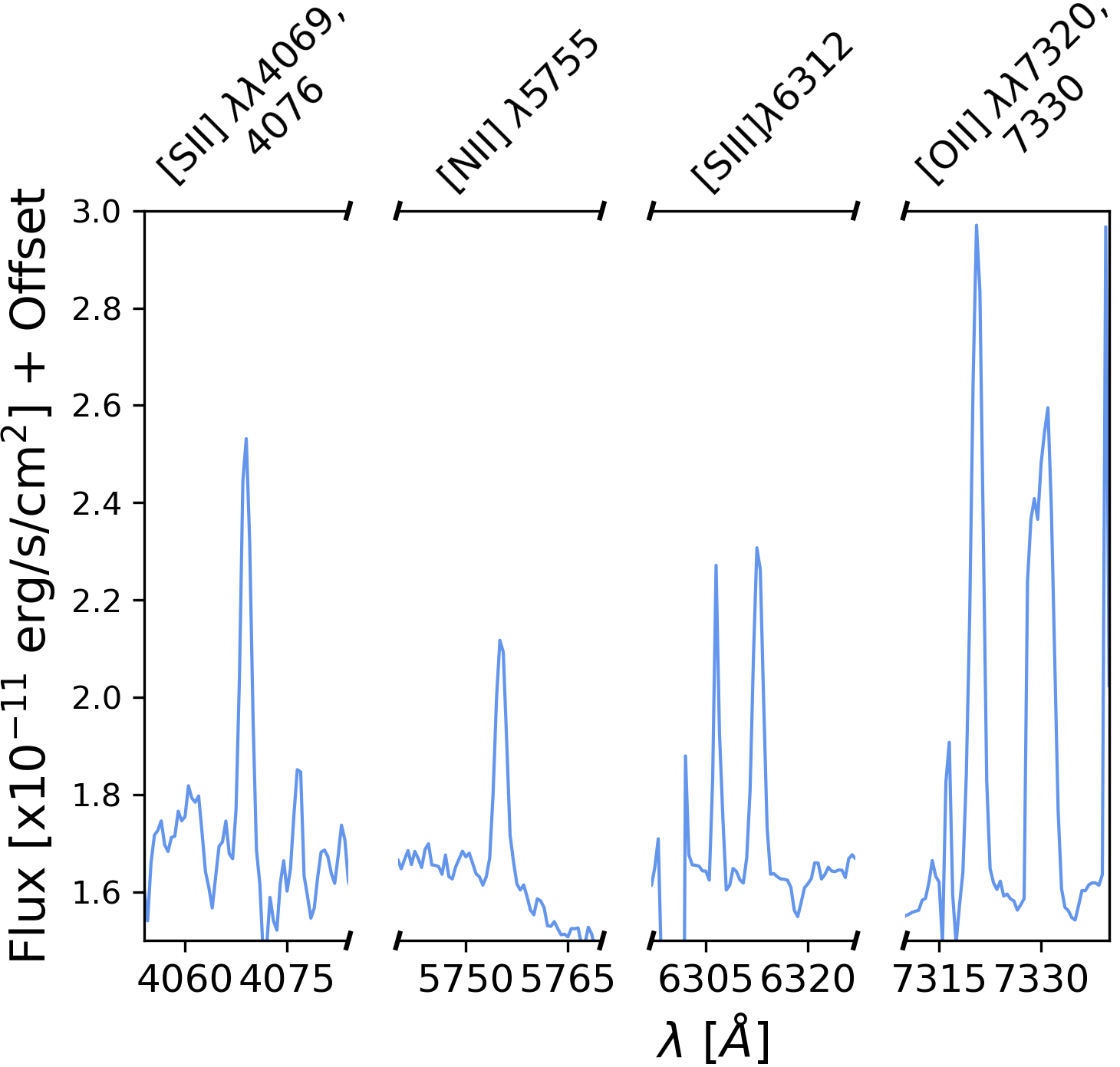}
        \end{minipage}
    \end{minipage}
    \caption{\textbf{Top:} Integrated spectrum of the full LVM pointing covering the Trifid Nebula.\\
    \textbf{Bottom:} A zoom-in collection of strong lines (left in magenta) and weak auroral lines (right in blue) taken from the above spectrum, each in a window of 30~\AA. The H$\alpha$ emission line is not shown here, as its height would dominate over all other lines.}
    \label{fig:spectrum_strongweak}
\end{figure*}

%%%%%%%%%%%%%%%%%%%%%%%%%%%%%%%%%%%%%%%%%%%%%%%%%%%%%%%%%%%%%%%%%%%%%%%%%%%%%%%%%%%%%%%%%%%%%%%%%%%%%%%%%%%%%%%%%%%%%%%

\section{Observations and Data Reduction}
\label{sec:observations}

\begin{figure}
    \centering
    \includegraphics[width=1\columnwidth]{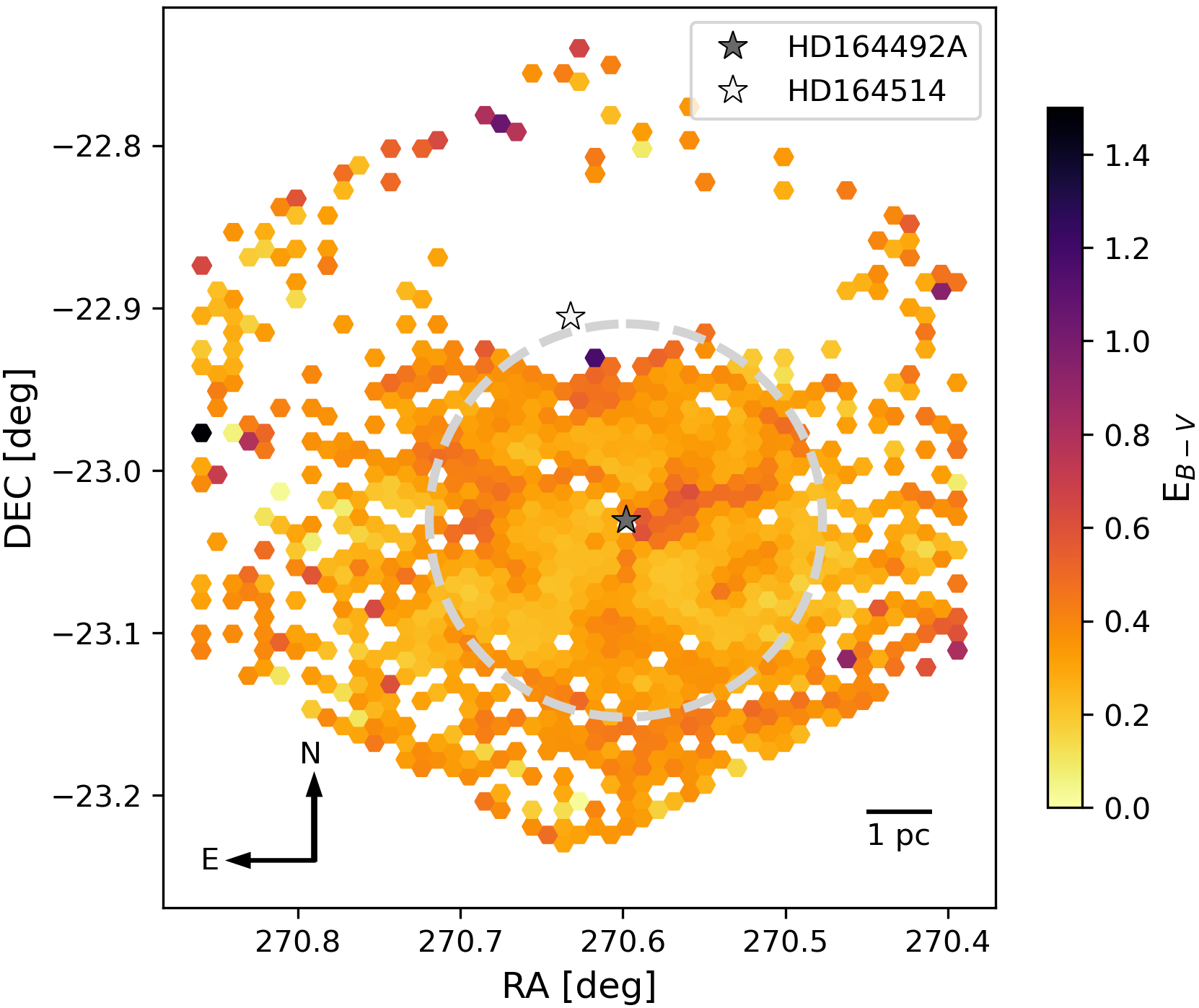}
    \caption{Spatial map of the dust extinction in M\,20.
    The position of HD\,164492A is marked with a grey star.
    The extent of the measured Strömgren sphere (Sect.~\ref{sec:Q0}) is shown by a grey dashed circle.
    The white regions arise from the limited sensitivity to the faint hydrogen emission lines in these spaxels and were masked (see Sect.~\ref{sec:observations}).
    A uncertainty map is shown in Fig.~\ref{fig:reddening_error}.} 
   % The density, temperature, and abundance determinations presented in this work do not extend to these northern positions and are therefore unaffected.}
    \label{fig:reddening}
\end{figure}

M\,20 was observed as part of the SDSS-V \citep{Kollmeier2019, Kollmeier2025} using the LVM, which hosts a stable wide field integral field unit to survey the ionized gas in the Milky Way, Magellanic clouds, and other Local Group galaxies \citep{Drory2024}.
The LVM is located at the Las Campanas Observatory in Chile's Atacama Desert and consists of four 16\,cm telescopes, each equipped with a fiber bundle (each fiber has a diameter of 35.3\,arcsec) that is fed into three DESI spectrographs with R$\sim$4000 across 3600-9800\,\AA\ \citep{Perruchot2018, Konidaris2020, Herbst2024}.
This allows us to detect many faint auroral lines, which are important for the measurements of physical conditions.
One of those four telescopes, hosting 1801 fibers with a total fill factor of 83\,\%, is used for the observation of the science targets, while the other three telescopes simultaneously observe spectrophotometric stars and sky fields, which are later used in the data reduction.
With this setup, one pointing of LVM 
covers a hexagon of 0.5$^{\circ}$ diameter, enabling us to observe M\,20 within only one LVM pointing, an angular resolution of 35.3\,arcsec,
and, in particular, a spatial resolution of 0.24~pc.
An overview of the full observing strategy will be given in \citet{Johnston2025}.

% this study
In this study, we use eight single-exposure (not dithered) early LVM science frames (taken during September 27-30, 2023), each with an integration time of 15 minutes, leading to a total of 2 hours and therefore a higher S/N (Signal-to-Noise) ratio than a single frame in LVM's survey mode.
The frames are reduced using version 1.1.1 of the dedicated LVM Data Reduction Pipeline (DRP) \citep{Mejia2024_DRP}, which removes instrumental features, sky emission, and calibrates the flux using individual stars in the science field.
Then, the eight reduced frames are combined into a single file.
This combination of spectra from multiple exposures is done by collecting the single fiber spectra, applying heliocentric velocity corrections, and merging spectra from fibers with matching positions using \textsc{astropy} \textit{sigma\_clip} (with $\sigma$~=~2) into a mean spectrum for each fiber.
The output single frame includes all key spectral extensions (e.g., flux and flux error) and a unified fiber position table for analysis.
With this, we reach a mean 3$\sigma$ H$\alpha$ sensitivity of around 6~$\times$~10$^{-16}$~erg/s/cm$^2$/arcsec$^2$ for the full frame.
A spectrum of the integrated LVM frame can be seen in Fig.~\ref{fig:spectrum_strongweak} together with a collection of zoom-in views of strong lines and weak auroral lines.

For each fiber, we fit the emission lines (Hydrogen recombination lines and collisionally excited lines, including also faint auroral lines: \oii$\lambda$$\lambda$3727,29; \sii$\lambda$4069; \oiii$\lambda$5007; \nii$\lambda$5755; \siii$\lambda$6312;  \nii$\lambda$6584; \sii$\lambda$$\lambda$6717,31; \oii$\lambda$7320, \oii$\lambda$7330; \siii$\lambda$9531) and the continuum spectrum independently and simultaneously using Gaussian profiles for the former and a linear profile for the latter.
This nearly linear continuum emission arises primarily from hydrogen free–free and free–bound emission and is removed from the line fits.
The \oii$\lambda\lambda$3727,29 doublet was fitted using a double Gaussian profile, deblending the two emission lines.

% quality cuts
To ensure good fits, all spaxels with S/N < 3 in the used diagnostic emission line and errors larger than the measured value itself were masked from the resulting maps (see Sect.~\ref{sec:results}).
Bright stars are negligible in the nebular spectra, as the large field-of-view of each LVM fiber means that individual stars contribute only to a small fraction of the total light collected, and thus do not dominate the signal in their corresponding fibers.

% general issues, be cautious
We note that our emission line measurements account for all known limitations of the current DRP version, such as sky subtraction, flat field correction, and flux calibration, since most of the selected emission lines, except He$\lambda$7281 and  \sii$\lambda$4069, are not affected by the sky subtraction, and the qualitative comparison of nearby emission lines would not be distorted with other calibrations.
Nevertheless, using a future version of the DRP might slightly change the absolute values of electron density, temperature, and ionic abundances by a few percent, but the measured gradients should remain unaffected.

%%%%%%%%%%%%%%%%%%%%%%%%%%%%%%%%%%%%%%%%%%%%%%%%%%%%%%%%%%%%%%%%%%%%%%%%%%%%%%%%%%%%%%%%%%%%%%%%%%%%%%%%%%%%%%%%%%%%%%%

\section{Analysis}
\label{sec:analysis}

%%%%%%%%%%%%%%%%%%%%%%%%%%%%%%%%%%%%%%%%%%%%%%%%%%%%%%%%%%%%%%%%%%%%%%%%%%%%%%%%%%%%%%%%%%%%%%%%%%%%%%%%%%%%%%%%%%%%%%%

\subsection{Reddening Correction}
\label{subsec:reddening}

\begin{figure*}
\begin{minipage}[l]{0.55\textwidth}
    \centering
    \includegraphics[width=\linewidth]{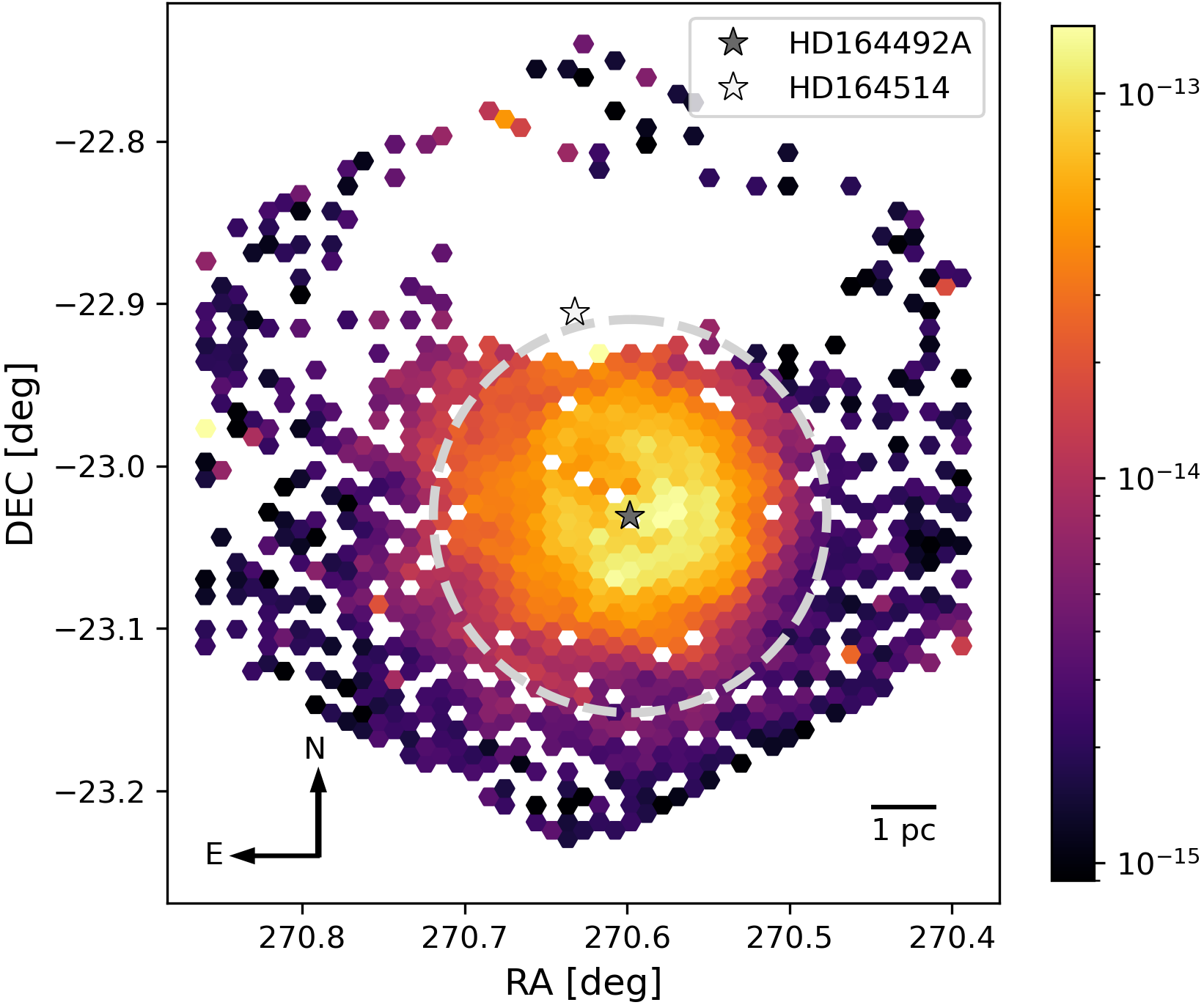}
\end{minipage}
\begin{minipage}[r]{0.45\textwidth}
    \centering
    \includegraphics[width=\linewidth]{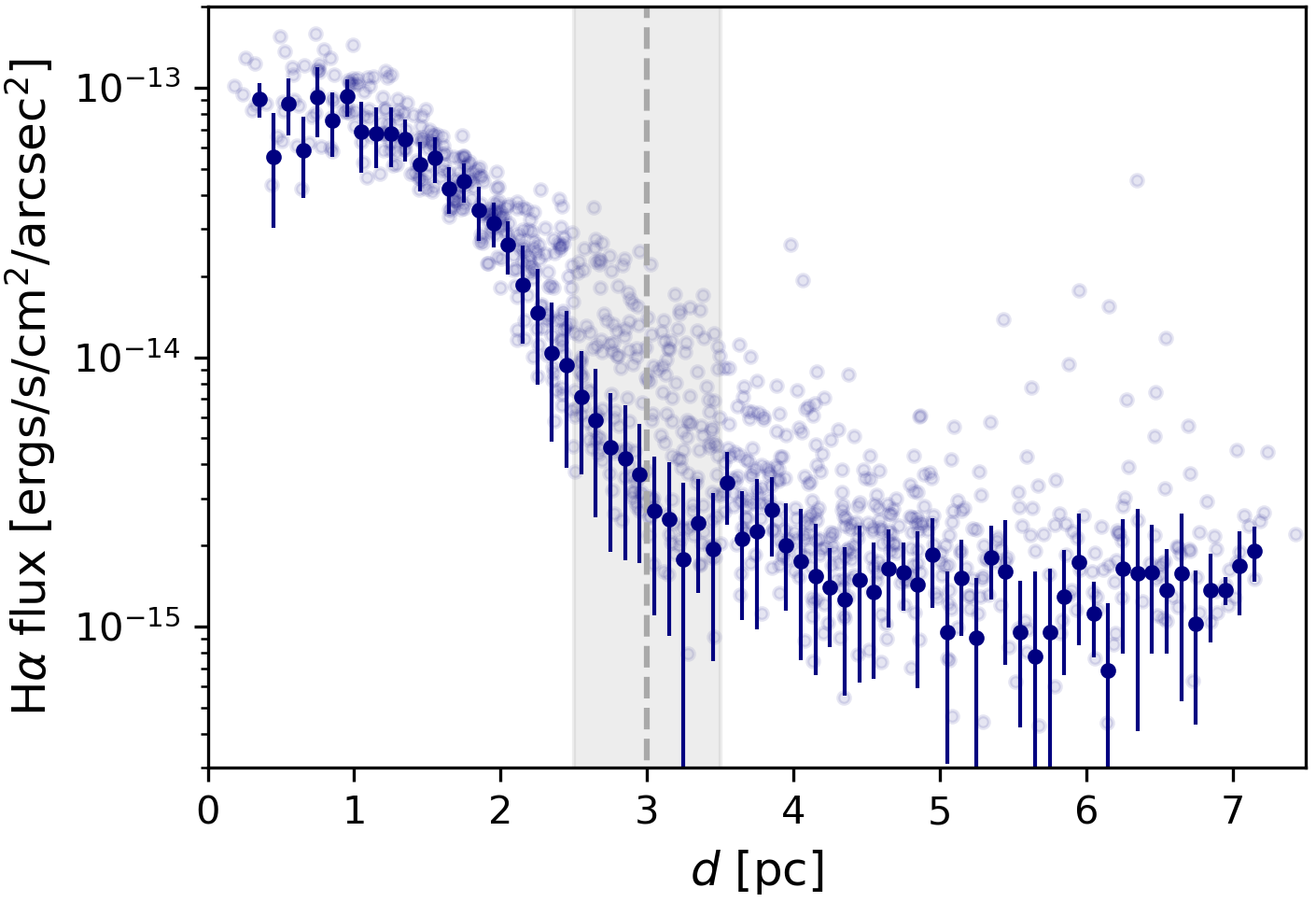}
\end{minipage}
\caption{\textbf{Left:} Map of the reddening corrected H$\alpha$ line flux as described in Sect.~\ref{subsec:reddening}. 
The position of HD\,164492A is marked with a grey star, while the position of HD\,164514 is marked with a white star.
The extent of the measured Strömgren sphere (Sect.~\ref{sec:Q0}) is shown by a grey dashed circle.\\
\textbf{Right:} Radial variation of the reddening corrected H$\alpha$ line flux as a function of the distance to the ionizing star HD\,164492A. 
Faint points represent individual spaxels, while opaque points show the uncertainty-weighted average of all spaxels within 0.1 pc wide distance bins.
Standard deviations are shown as error bars.
The extent of the measured Strömgren sphere (Sect.~\ref{sec:Q0}) is shown by a grey dashed line together with the error as a grey shaded region.}
\label{fig:Halpha}
\end{figure*}

We apply a reddening correction to the measured fluxes of all spaxels to account for extinction.
To do this, we use \textsc{pyneb} v.1.1.19b2 \citep{Luridiana2015, Morisset2020} (together with the atomic data set used in \citealt{Mendez-Delgado2025}) which follows the well-known relationship:% Equation 1 from \citet{Lopez-Sanchez2006}:
\begin{equation}
    \text{c}\left(\text{H}\beta\right)\ \text{=}\ \frac{1}{f\left(\lambda\right)}\ \text{log}  \left(
     \frac{\frac{I\left(\lambda\right)}{I\left(\text{H}\beta\right)}}{\frac{F\left(\lambda\right)}{F\left(\text{H}\beta\right)}} \right)
    \label{eq:reddening}
\end{equation}
where \text{c}(\text{H}$\beta$) is the reddening coefficient, $f\left(\lambda\right)$ is the reddening law normalized to H$\beta$, $I\left(\lambda\right)$ is the theoretical flux and $F\left(\lambda\right)$ is the observed (uncorrected) flux.

To calculate the reddening correction factor \text{c}(\text{H}$\beta$) for our \hii\ region, we compare the measured ratios of hydrogen Balmer and Paschen lines with the theoretical values provided by \citet{Storey1995} at an electron temperature of 10000~K and electron density of 100~cm$^{-3}$.
Using other reasonable temperature or density combinations (e.g., the ones calculated in this paper) does not significantly affect the reddening correction factor or the final results and conclusions.
We use the reddening law of \citet{Cardelli1989} modified by \citet{Blagrave2007} together with an $R_V$~=~5.5 as measured by \citet{greve2010} and \citet{Cambresy2011}.

Using \textsc{pyneb}, the reddening correction factor of each ratio is then extracted with the \textit{cHbeta} function.
To get a robust measure, we adopt as our final value of \text{c}(\text{H}$\beta$) the median of all calculated correction factors derived from the various hydrogen line ratios.
Here, we make sure to only use hydrogen line ratios that are well measured and yield consistent correction factors 
(H$\lambda$3771/H$\lambda$6563, 
H$\lambda$3750/H$\lambda$9229, 
H$\lambda$3771/H$\lambda$9229, 
H$\lambda$3835/H$\lambda$9229, 
H$\lambda$4102/H$\lambda$9229, 
H$\lambda$4340/H$\lambda$9229, 
H$\lambda$4861/H$\lambda$9229, 
H$\lambda$6563/H$\lambda$9229), to exclude ratios affected by flux calibration issues.
Finally, each measured emission line flux is corrected for extinction using Eq.~\eqref{eq:reddening}
%\begin{equation}
%    \frac{I\left(\lambda\right)}{I\left(\text{H}\beta\right)}\ \text{=}\ \frac{F\left(\lambda\right)}{F\left(\text{H}\beta\right)} \times 10^{\text{c}\left(\text{H}\beta\right)\,f\left(\lambda\right)}
%    \label{eq:reddening_applied}
%\end{equation}
, where the correction term is acquired with \textsc{pyneb}'s \textit{getCorr} function \citep{Luridiana2015}.
The $E_{B-V}$ as shown in Fig.~\ref{fig:reddening} is related to \text{c}(\text{H}$\beta$) as:
\begin{equation}
    \left(1-f\left(\lambda\right)\right)\ \text{c}\left(\text{H}\beta\right)\ \text{=}\ 0.4\ E_{B-V}\ X\left(\lambda\right)
    \label{eq:EBV}
\end{equation}
where $X\left(\lambda\right)\ \text{=}\ A\left(\lambda\right)\,\text{/}\,E_{B-V}\ \text{=}\ R_V A\left(\lambda\right)\,\text{/}\,A_V$.
The mean $E_{B-V}$ for the resolved view is 0.34~$\pm$~0.01.
A map of the reddening corrected H$\alpha$ flux together with its radial profile can be seen in Fig.~\ref{fig:Halpha}.

%%%%%%%%%%%%%%%%%%%%%%%%%%%%%%%%%%%%%%%%%%%%%%%%%%%%%%%%%%%%%%%%%%%%%%%%%%%%%%%%%%%%%%%%%%%%%%%%%%%%%%%%%%%%%%%%%%%%%%%

\subsection{Calculation of Physical Properties}
\label{subsec:properties}

%%%%%%%%%%%%%%%%%%%%%%%%%%%%%%%%%%%%%%%%%%%%%%%%%%%%%%%%%%%%%%%%%%%%%%%%%%%%%%%%%%%%%%%%%%%%%%%%%%%%%%%%%%%%%%%%%%%%%%%

\subsubsection{Electron Densities}
\label{analysis:densities}

The calculation of electron density, temperature, and ionic abundances was also done with the \textsc{pyneb} package.

% CEL
To measure the resolved electron density, we used the common ratios of the two CEL line doublets: \sii$\lambda$6717/$\lambda$6731 and \oii$\lambda$3727/$\lambda$3729, together with the \textsc{pyneb} \textit{getTemDen} function. 
An initial guess of 10000~K for the electron temperature is used in this step.
To get a stable measurement for the densities and associated uncertainties of our measurement, we executed 1000 Monte Carlo (MC) realizations of these calculations, 
adding random noise to the initial flux values in each realization. 
This random noise ranges between zero and the uncertainty of the initial flux measurements.
The final density values were taken as the median of the 1000 MC results, whereas the uncertainties result from the MC standard deviation.
Because this process would result in a huge amount of computing time when done for each spaxel, the machine learning tool \textsc{ai4neb}\footnote{\href{https://github.com/Morisset/AI4neb}{https://github.com/Morisset/AI4neb}} was used to accelerate the computations.
In this approach, a neural network is first trained on examples of line ratios together with their corresponding T$_e$ and n$_e$ values, and is then applied to predict the physical parameters directly from the observations.

%%%%%%%%%%%%%%%%%%%%%%%%%%%%%%%%%%%%%%%%%%%%%%%%%%%%%%%%%%%%%%%%%%%%%%%%%%%%%%%%%%%%%%%%%%%%%%%%%%%%%%%%%%%%%%%%%%%%%%%

\subsubsection{Electron Temperatures}
\label{analysis:temperature}

% temperature
For the measurement of the resolved electron temperatures, we used the common ratios of temperature-sensitive lines: \nii$\lambda$6584/$\lambda$5755, \oii$\lambda$$\lambda$(3727+3729)/$\lambda$7320\footnote{we did not include $\lambda$7330, as it blends with an OH airglow sky line producing emission around $\lambda$7330 evident from the double peak and broader shape compared to $\lambda$7320 in Fig.~\ref{fig:spectrum_strongweak}}, \sii$\lambda$$\lambda$(6717+6731)/$\lambda$4069 and \siii$\lambda$9531/$\lambda$6312. 
In this step, we used \textsc{pyneb}'s \textit{getTemDen} function with a density value taken to be the mean of our two density measurements, n$_e$(\oii) and n$_e$(\sii).
Note that using either n$_e$(\oii), n$_e$(\sii), or the mean value does not significantly affect the resulting electron temperatures and abundances.
Similar to the density calculations, 1000 MC realizations were executed using \textsc{ai4neb} and the final temperature measurements were calculated as the median of those realizations.

% possible overestimation?
The measurement of T$_e$(\sii) needs to be handled with caution as there might be inaccurate or excessive sky subtraction in the \sii$\lambda$4069 line, potentially leading to an overestimation of the temperatures.
Because of this, we do not use this diagnostic in further calculations of the abundances.
% more calculations
We also attempted to calculate the electron temperatures using the [Ar~\textsc{iii}]$\lambda$7135/$\lambda$5192 and \oiii$\lambda$5007/$\lambda$4363 ratios.
However, the fainter emission lines that are required for these diagnostics (e.g., [Ar~\textsc{iii}]$\lambda$5192, \oiii$\lambda$4363) were not measured with S/N~>~3 in a sufficient number of fibers.

%%%%%%%%%%%%%%%%%%%%%%%%%%%%%%%%%%%%%%%%%%%%%%%%%%%%%%%%%%%%%%%%%%%%%%%%%%%%%%%%%%%%%%%%%%%%%%%%%%%%%%%%%%%%%%%%%%%%%%%

\subsubsection{Abundances}
\label{analysis:abundances}

% abundance
To measure the resolved ionic abundances of oxygen, we used \textsc{pyneb}'s \textit{getIonAbundance} function together with the mean electron density value.
For the calculation of the O$^{+}$ abundance, we used the \oii$\lambda$$\lambda$3727+3729 lines together with the mean temperature of the low-ionization states T$_e$(\nii) and T$_e$(\oii).
We excluded the T$_e$(\sii) measurement from this calculation, as its values might be overestimated (see Sect.~\ref{subsec:temperature}).
For the calculation of the O$^{2+}$ abundance, we used the \oiii$\lambda$5007 line together with the electron temperature for high-ionization states, T$_e$(\siii).
We could not use the temperature of \oiii\ as the $\lambda$4363 line is not detected.
As before, the use of different density or temperature diagnostics for the abundance determination does not significantly affect the final abundance values.
A similar MC run as above was carried out for these measurements, with a lower number of 100 realizations, as the abundance calculations take a much longer time compared to the calculations of densities and temperatures.

%%%%%%%%%%%%%%%%%%%%%%%%%%%%%%%%%%%%%%%%%%%%%%%%%%%%%%%%%%%%%%%%%%%%%%%%%%%%%%%%%%%%%%%%%%%%%%%%%%%%%%%%%%%%%%%%%%%%%%%

\subsection{Integrated Measurements}
\label{subsec:integrated_calc}

% integrated spectra
In order to quantify the differences between spatially resolved and unresolved measurements of electron density, temperature, and oxygen abundance, and constrain the impact of inhomogeneities, we construct an integrated spectrum by co-adding all fiber spectra from the full LVM pointing into a single spectrum on which the emission line analysis is performed.
Although one could restrict the integration to the fibers located within a sphere around HD\,164492A, we adopt the full tile integration to better mimic extragalactic observations, where individual \hii\ regions cannot typically be resolved and the integrated spectrum naturally includes surrounding diffuse emission and unrelated structures. 
We also verified that restricting the integration to a sphere with 2.43~pc radius (the theoretical Strömgren radius of an O\,7.5\,V star) does not significantly alter the derived physical parameters.
A collection of strong and weak lines observed with the full integrated spectrum is shown in Fig.~\ref{fig:spectrum_strongweak}, and the measured line fluxes can be found in App.~\ref{app:fluxes}.

The same analysis as described in Sect.~\ref{subsec:reddening} and \ref{subsec:properties} is then repeated on the integrated spectrum.
Using the same hydrogen line ratios as for the resolved measurements, we get $E_{B-V}$~=~0.36~$\pm$~0.01.
% more temperature measurements
The improved S/N of the integrated spectra also allows us to measure the electron temperature from the ratio of the He~\textsc{i}~$\lambda$7281/$\lambda$6678 recombination lines \citep{Zhang2005, Mendez-Delgado2025}.
Because this is not a standard CEL-diagnostic in \textsc{pyneb}, we define a function that takes the observed line ratio and returns the corresponding electron temperature.
For this, we calculate the theoretical emissivity ratios for various possible temperatures in the range of 5000 to 20000~K and use \textsc{scipy}'s \textit{interp1d} to construct an interpolation function that lets us constrain T$_e$(He~\textsc{i}).
As with previous measurements, we repeat the calculation of the helium electron temperature for 1000 MC realizations, take the mean value as our final result, and the standard deviation as an uncertainty estimate.

\begin{figure*}
    \centering
    \includegraphics[width=0.85\linewidth]{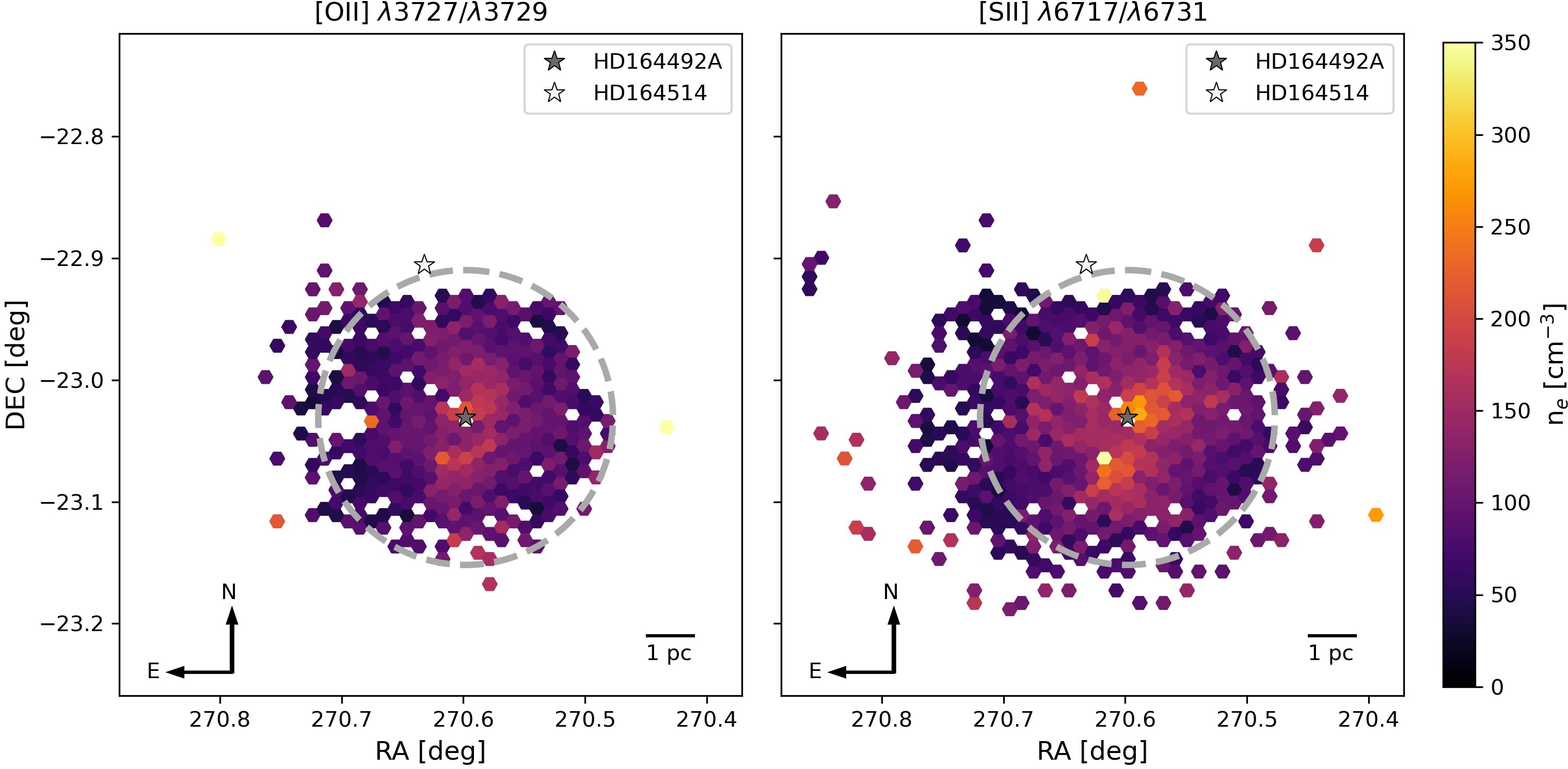}
    \caption{Maps of the electron densities. 
    In all maps, the position of HD\,164492A is marked with a grey star, while the position of HD\,164514 is marked with a white star.
    The extent of the measured Strömgren sphere (Sect.~\ref{sec:Q0}) is shown by a grey dashed circle.
    Uncertainty maps corresponding to each density tracer are shown in Fig.~\ref{fig:density_error}.}
    \label{fig:density}
\end{figure*}
\begin{figure}
    \centering
    \includegraphics[width=0.98\columnwidth]{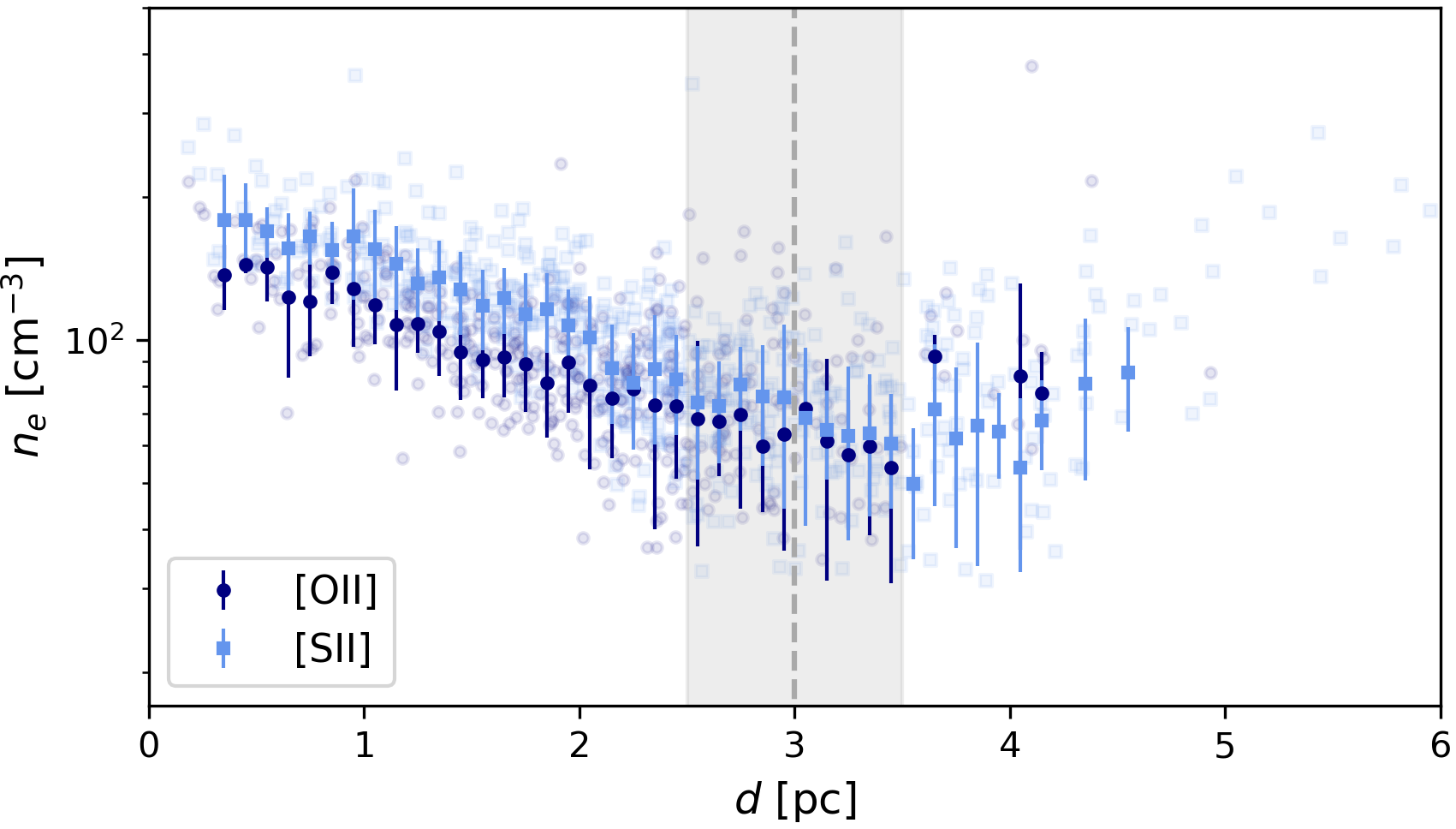}
    \caption{Radial variation of the electron densities (n$_e$(\oii) in dark blue circles and n$_e$(\sii) in light blue squares) as a function of the distance to the ionizing star HD\,164492A. 
    Faint points represent individual spaxels, while opaque points show the uncertainty-weighted average of all spaxels within 0.1 pc wide distance bins.
    Standard deviations are shown as error bars.
   The extent of the measured Strömgren sphere (Sect.~\ref{sec:Q0}) is shown by a grey dashed line together with the error as a grey shaded region.}
    \label{fig:density_radial}
\end{figure}

%%%%%%%%%%%%%%%%%%%%%%%%%%%%%%%%%%%%%%%%%%%%%%%%%%%%%%%%%%%%%%%%%%%%%%%%%%%%%%%%%%%%%%%%%%%%%%%%%%%%%%%%%%%%%%%%%%%%%%%

\subsection{Estimating the Stellar Ionizing Rate $Q_0$}
\label{sec:Q0}

Because the Trifid Nebula is only ionized by a single O star HD\,164492A, it shows a rather simple ionization structure in which we can probe spherical symmetries.
The H$\alpha$ flux as seen in Fig.~\ref{fig:Halpha} shows azimuthal variations smaller than 10~\%.

The ionization equilibrium implies that, for a filled sphere of gas, the rate of photoionizations is balanced by the rate of recombinations:
\begin{equation}
Q_0 [s^{-1}] = \frac{4}{3}\ \pi\ \text{R}^3\ f_{\text{fill}}\ \text{n}_e^2\ \frac{\alpha_B}{f_{H^+}}
\label{eq:Q0}
\end{equation}
\citep{Osterbrock1974, Draine2011}, where $Q_0$ is the number of ionizing photons emitted by the central source, R is the size of the considered sphere (it may be the Strömgren radius if the region is radiation-bounded), $f_{\text{fill}}$ is the filling factor (or clumpiness factor), n$_e$ is the electron density, $\alpha_B$ is the case B recombination coefficient, and $f_{H^+}$ is the fraction of ionizing photons that actually ionize H (and not dust or other elements; the covering factor $\Omega/4\pi$ also enters in $f_{H^+}$, as well as the escaping fraction due to a matter-bounded nebula). 
The case B recombination coefficient can be expressed by: 
\begin{equation}
\alpha_B [\text{cm}^3\ \text{s}^{-1}] = 2.59 \times 10^{-13}\ \text{T}_4^{-0.833 - 0.034\ \text{ln}(\text{T}_4)}
\label{eq:alpha}
\end{equation}
following Tab.~14.1 from \citet{Draine2011}, where T$_4$~=~T$_e/10^4$K. 

The value of $Q_0$ can then be estimated from the observations presented in this work. From the resolved H$\alpha$ flux in Fig.~\ref{fig:Halpha} (right), one can see that the flux is high in the central region and decreases heavily until a distance of around 3.0~pc. 
This sets our observed radius to R$_{\text{obs}}$~=~3.0~$\pm$~0.5~pc.
The main difficulties are to determine the filling factor $f_{\text{fill}}$ and the electron density n$_e$. 
The latest derived from the \sii$\lambda$6717/$\lambda$6731 line ratio seems to decrease with the distance to the center, as shown in Fig.~\ref{fig:density_radial} described below. 
This only gives insights into the low ionization region density; the inner, higher ionization part of the nebula may be at higher densities. 
A way to remove these difficulties is to consider the integrated flux H$\alpha$, which can be expressed as:
\begin{equation}
L\left(\text{H}\alpha\right)\ [\text{erg\ s}^{-1}] = \frac{4}{3}\ \pi\ \text{R}^3\ f_{\text{fill}}\ \text{n}_e^2\ \epsilon_\alpha 
\label{eq:LHa}
\end{equation}
where $\epsilon_\alpha$, the emissivity of H$\alpha$, is taken from Eq.~(14.8) of \citet{Draine2011}: 
\begin{equation}
\begin{split}
\epsilon_\alpha\ [\text{erg\ s}^{-1}\ \text{cm}^3] &= h\ \nu_\alpha\ \alpha_{\text{eff, H}\alpha} \\
&= 3.55 \times 10^{-25}\ \text{T}_4^{-0.942-0.031\ \text{ln(T}_4)}
\end{split}
\label{eq:epsilon}
\end{equation}
Together with Eqs.~\eqref{eq:Q0} and \eqref{eq:LHa}, this leads to the following relation:
\begin{equation}
\begin{split}
Q_0\ [\text{s}^{-1}] &= L\left(\text{H}\alpha\right)\ \frac{\alpha_B(\text{T}_e)}{\epsilon_\alpha(\text{T}_e)\ f_{H^+}} \\
&= 7.32 \times 10^{11} [\text{erg}^{-1}]\ L\left(\text{H}\alpha\right)\ [\text{erg\ s}^{-1}]\  \frac{\text{T}_4^{0.1}}{f_{H^+}}
\end{split}
\label{eq:Q0_2}
\end{equation}
where for T$_e$~=~10000~K, the exponent of T$_4$ simplifies to 0.1.
The only parameter to be determined remains $f_{H^+}$. 

The integrated spectrum over all LVM spaxels allows us to define $L\left(\text{H}\alpha\right)$.
We use the distance to the Trifid Nebula of $d$~=1.42~$^{+0.09}_{-0.08}$~kpc (see Sect.~\ref{sec:introduction}) to calculate from the observed integrated and reddening corrected H$\alpha$ flux $F\left(\text{H}\alpha\right)$ the H$\alpha$ luminosity $L\left(\text{H}\alpha\right)$ using:
\begin{equation}
\begin{split}
L\left(\text{H}\alpha\right)\ [\text{erg\ s}^{-1}]\ \text{=}\ 4\pi \left(d\,\text{[cm]}\right)^2\ F\left(\text{H}\alpha\right) \,[\text{erg\ s$^{-1}$\ cm$^{-2}$}]\ \\
\times \frac{1}{f_{\text{LVM}}}
\label{eq:luminosity}
\end{split}
\end{equation}
where $f_{\text{LVM}}$ is the fill factor of the LVM fibers of 0.83 as mentioned in Sect.~\ref{sec:observations}. 
The value we obtain is $L\left(\text{H}\alpha\right)$~=~(8.66~$\pm$~0.5)~$\times$~10$^{36}$~erg/s.

Using Eq.~\eqref{eq:Q0_2} in the case of a pure hydrogen nebula of constant n$_e$~=~95~cm$^{-3}$ with $f_{H^+}~=~1$ and at T$_e~=~8600$K, we obtain $Q_0~=~(6.25~\pm~0.36)~\times~10^{48}\ \text{s}^{-1}$, and with Eq.~\eqref{eq:Q0} and $f_{\text{fill}}~=~1$ a radius of R~=~2.7~$\pm$~0.3~pc. 
A filling factor of $f_{\text{fill}} = 0.7$ is needed to recover a Strömgren radius of 3.0~pc.

To determine a value of $f_{H^+}$ more tailored to the Trifid conditions, we run a photoionization toy model using the \textsc{cloudy} code \citep[v25.00, see][]{2025Gunasekera_arXi}, applying a more realistic ionizing flux distribution, a decreasing electron density law, dust mixed with the ionized gas, and a filling factor $f_{\text{fill}}~=~0.73$ (computational details can be found in App.~\ref{app:toymodel}). 
The resulting value for $f_{H^+}$ is 0.35, leading to $Q_0 = 2.09 \times 10^{12}\  L\left(\text{H}\alpha\right) = (1.8~\pm~0.1) \times 10^{49}$~s$^{-1}$. 
This must be considered as a lower value, as the model describes a radiation-bounded nebula, with a complete covering factor.

\begin{figure*}
    \centering
    \includegraphics[width=1\linewidth]{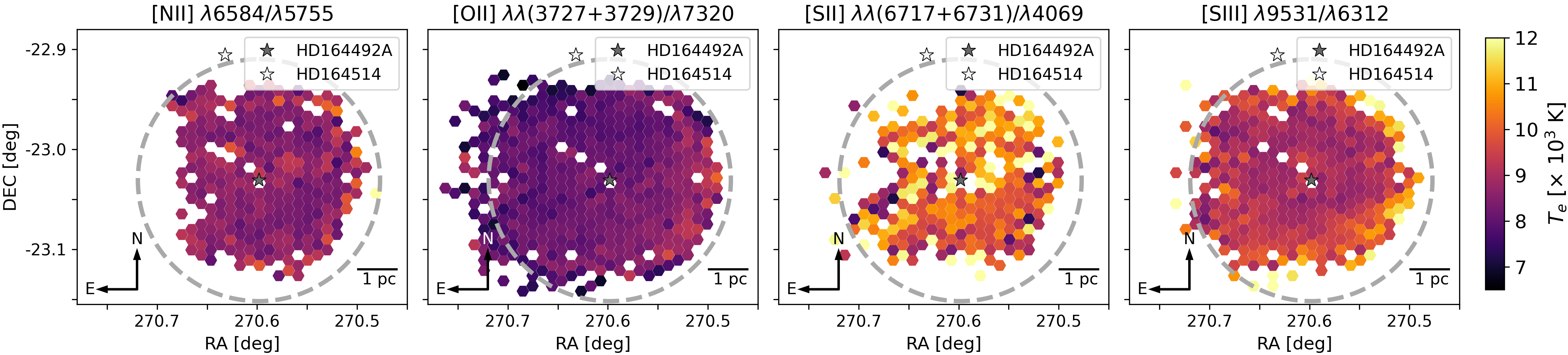}
    \vspace{-0.5cm}
    \caption{Maps of electron temperatures from different diagnostics.
    In all maps, the position of HD\,164492A is marked with a grey star, while the position of HD\,164514 is marked with a white star.
    The extent of the measured Strömgren sphere (Sect.~\ref{sec:Q0}) is shown by a grey dashed circle.
    Uncertainty maps corresponding to each temperature tracer are shown in Fig.~\ref{fig:temperature_error}.}
    \label{fig:temperature}
\end{figure*}

\begin{figure}
    \centering
    \includegraphics[width=0.9\columnwidth]{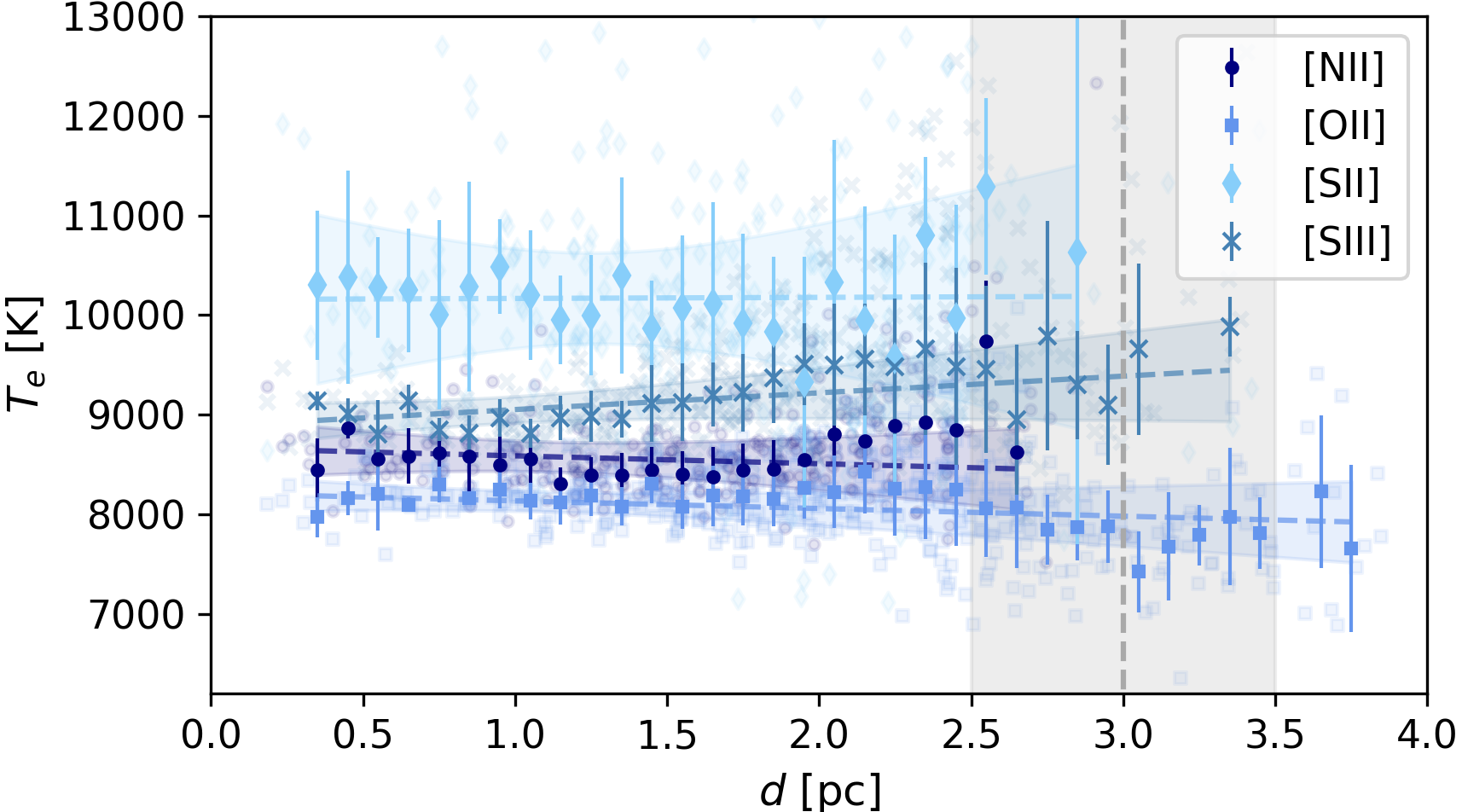}
    \caption{Radial variation of the electron temperatures (T$_e$(\nii) in dark blue circles, T$_e$(\oii) in blue squares, T$_e$(\sii) in light blue diamonds and T$_e$(\siii) in blue crosses) as a function of the distance to the ionizing star HD\,164492A. 
    Faint points represent individual spaxels, while opaque points show the uncertainty-weighted average of all spaxels within 0.1 pc wide distance bins.
    Standard deviations are shown as error bars.
    The extent of the measured Strömgren sphere (Sect.~\ref{sec:Q0}) is shown by a grey dashed line together with the error as a grey shaded region.
    The linear fits (slopes can be seen in Eq.~\eqref{eq:temperatures}) to the single temperatures are displayed as dashed lines together with the 3$\sigma$ uncertainties as shaded regions in the corresponding colors.}
    \label{fig:temperature_radial}
\end{figure}

%%%%%%%%%%%%%%%%%%%%%%%%%%%%%%%%%%%%%%%%%%%%%%%%%%%%%%%%%%%%%%%%%%%%%%%%%%%%%%%%%%%%%%%%%%%%%%%%%%%%%%%%%%%%%%%%%%%%%%%

\section{Results}
\label{sec:results}

%%%%%%%%%%%%%%%%%%%%%%%%%%%%%%%%%%%%%%%%%%%%%%%%%%%%%%%%%%%%%%%%%%%%%%%%%%%%%%%%%%%%%%%%%%%%%%%%%%%%%%%%%%%%%%%%%%%%%%%

\subsection{Spatially Resolved Electron Densities}
\label{subsec:density}

% overall structure CEL densities
The electron density maps in Fig.~\ref{fig:density} show a similar overall structure for both measured line ratios, with higher densities around 150~cm$^{-3}$ for n$_e$(\oii) to 200~cm$^{-3}$ for n$_e$(\sii) in the central region near the ionizing source. 
Both maps indicate a decrease in density with increasing distances from the central ionizing star HD\,164492A to a minimum of around 30~cm$^{-3}$ for n$_e$(\oii) and 40~cm$^{-3}$ for n$_e$(\sii).
This negative radial gradient in density is also clearly visible in Fig.~\ref{fig:density_radial}, showing the radial density distribution, where the density minimum is reached at a distance of approximately 3-4~pc from the center.
It can also be seen that the electron density measured with the \oii\ doublet is systematically lower by approximately 10 to 50 cm$^{-3}$ than the n$_e$(\sii) measurement, which can be explained by the fact that the two diagnostics have different critical densities, leading to small systematic offsets.

At larger distances, the overall \sii\ density seems to increase again towards the outer edges of the LVM pointing, but these values also have uncertainties larger than 100~cm$^{-3}$ (see Fig.~\ref{fig:density_error}).
Because of a lower S/N in the \oii\ lines, we do not cover much of these fainter outer regions with the \oii\ density measurement. 
Thus, the apparent increase in density might be an artifact of large uncertainties in the outer regions, although we cannot exclude a contribution from the reflection nebula, which could affect the observed line ratios and mimic a density enhancement.

% single density enhancements
The radial density variation in Fig.~\ref{fig:density_radial} shows a small increase in densities around 1~pc away from the central star, for both diagnostics.
This increase is associated with a higher density concentration (at RA~$\approx$~270.61$^\circ$ and DEC~$\approx$~-23.07$^\circ$) visible south-east of HD\,164492A in Fig.~\ref{fig:density}.
Another region of enhanced electron density (at RA~$\approx$~270.58$^\circ$ and DEC~$\approx$~-23.02$^\circ$) can be seen north-west and very close to the ionizing star in Fig.~\ref{fig:density}, causing the highest density values of the most central part in Fig.~\ref{fig:density_radial}.

%%%%%%%%%%%%%%%%%%%%%%%%%%%%%%%%%%%%%%%%%%%%%%%%%%%%%%%%%%%%%%%%%%%%%%%%%%%%%%%%%%%%%%%%%%%%%%%%%%%%%%%%%%%%%%%%%%%%%%%

\subsection{Spatially Resolved Electron Temperatures}
\label{subsec:temperature}

\begin{figure*}
\begin{minipage}[l]{0.66\textwidth}
    \centering
    \includegraphics[width=\linewidth]{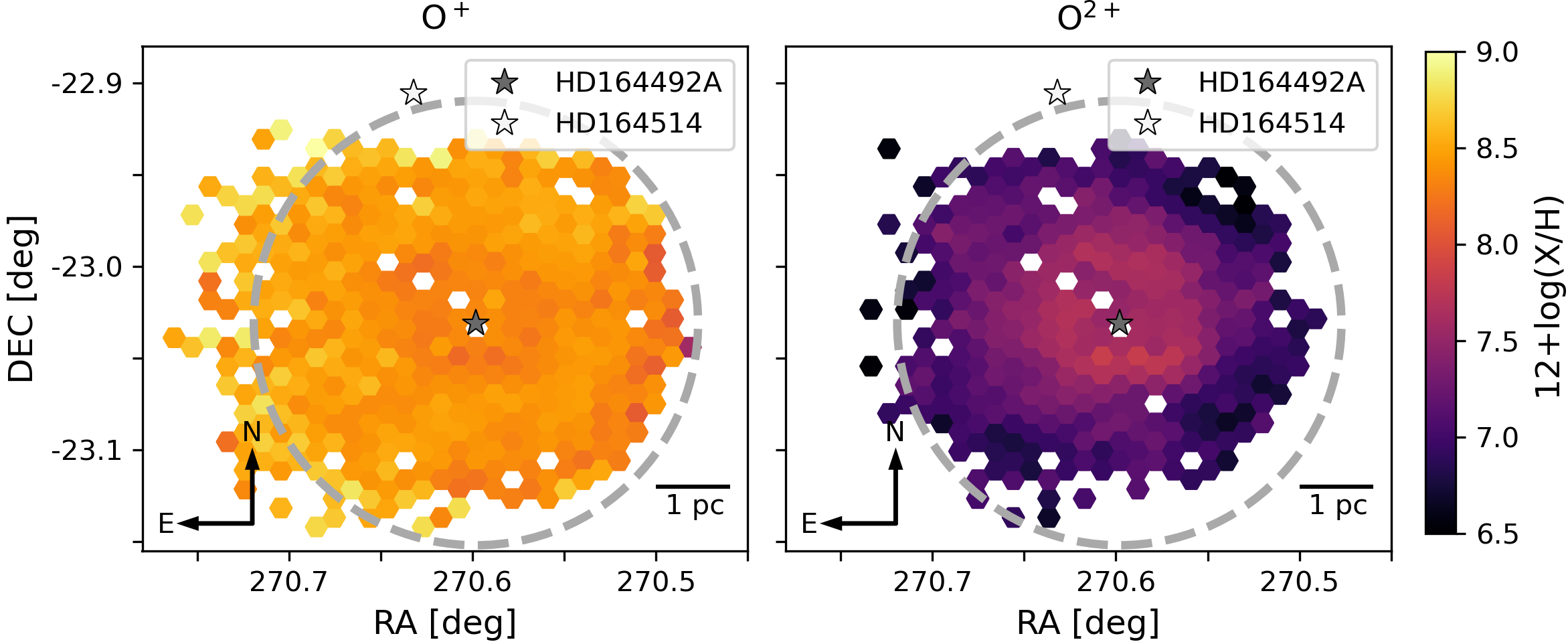}
\end{minipage}
\begin{minipage}[r]{0.335\textwidth}
    \centering
    \includegraphics[width=\linewidth]{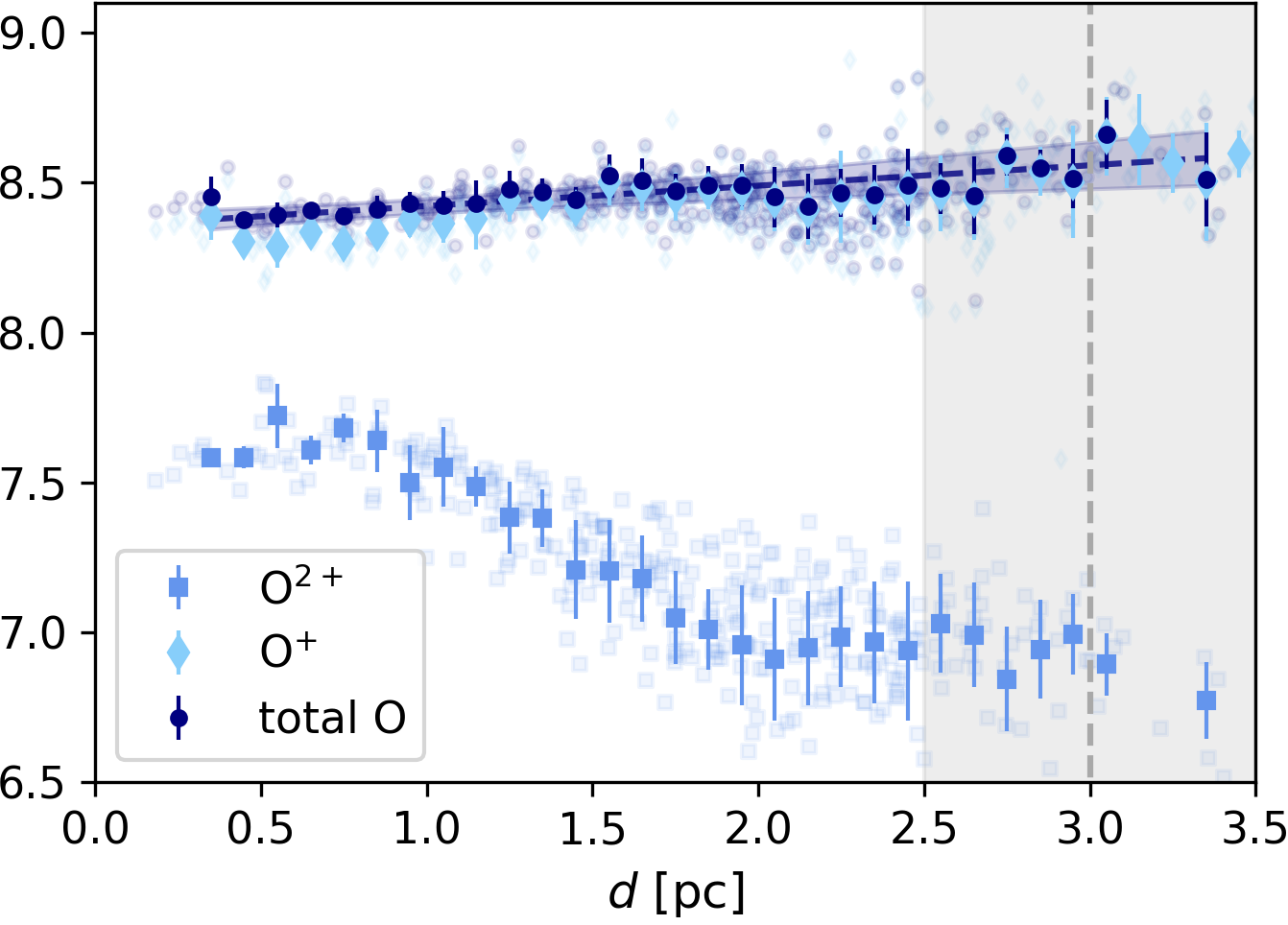}
\end{minipage}
\caption{\textbf{Left:} Maps of the O$^{+}$ (left) and O$^{2+}$ (right) abundances. In both maps, the position of HD\,164492A is marked with a grey star, while the position of HD\,164514 is marked with a white star.
The extent of the measured Strömgren sphere (Sect.~\ref{sec:Q0}) is shown by a grey dashed circle.
Uncertainty maps can be seen in Fig.~\ref{fig:Oabundance_error}.\\
\textbf{Right:} Radial variation of the oxygen abundances (O$^{+}$ in light blue diamonds, O$^{2+}$ in blue squares, and the total O abundance in dark blue dots) as a function of the distance to the ionizing star HD\,164492A. 
Faint points represent individual spaxels, while opaque points show the uncertainty-weighted average of all spaxels within 0.1 pc wide distance bins.
Standard deviations are shown as error bars.
The extent of the measured Strömgren sphere (Sect.~\ref{sec:Q0}) is shown by a grey dashed line together with the error as a grey shaded region.
The linear fit (slope can be seen in Eq.~\eqref{eq:abundance}) to the total oxygen abundance is displayed as a dashed line together with the 3$\sigma$ uncertainty as a shaded region.}
\label{fig:Oabundances}
\end{figure*}

% overall structure
Spatially resolved maps of the electron temperatures measured through different diagnostic line ratios can be seen in Fig.~\ref{fig:temperature}, while uncertainties of these maps can be seen in Fig.~\ref{fig:temperature_error}.
In most maps, no clear structures or radial gradients can be found; only T$_e$(\siii) might give rise to a positive radial temperature gradient, but the scatter and uncertainties across these outer edges are also large (>~1000~K, see Fig.~\ref{fig:temperature_error}).
On average, the measured T$_e$(\nii) and T$_e$(\oii) are lowest with mean values of 8541~$\pm$~382~K and 8180~$\pm$~324~K, respectively, while the T$_e$(\sii) values are highest with a mean of 10024~$\pm$~905~K.
The measured T$_e$(\siii) shows values in between the other diagnostics with a mean electron temperature of 9035~$\pm$~374~K.
All the mean values referenced here are reported in Tab.~\ref{tab:integrated_results}.
Missing spaxels in the maps, especially in the T$_e$(\sii) distribution, result from low S/N in the \sii$\lambda$4069 line, which leads to poor Gaussian fits that were subsequently discarded (see Sect.~\ref{sec:observations}). 
This effect occurs in all temperature maps but is most pronounced for \sii\ due to the weakness of the auroral line.

% homogeneous?
Looking at the radial electron temperature trends in Fig.~\ref{fig:temperature_radial}, the different temperature regimes of the diagnostics become even clearer, with T$_e$(\oii) < T$_e$(\nii) < T$_e$(\siii) < T$_e$(\sii).
To test their homogeneity, we also fitted linear functions to all temperatures with a least-squares algorithm (resulting gradients can be found in App.~\ref{app:gradients}).
However, considering the uncertainties per spaxel and the 3$\sigma$ confidence band from the fits (shaded regions in Fig.~\ref{fig:temperature_radial}) into account, which can be of the order of 1000~K, especially at the outer edges, there is consistency with a flat temperature distribution.
Also, azimuthal variations in electron temperatures are at maximum about the size of the measurement uncertainties.
Given that T$_e$(\sii) is systematically higher than the other diagnostics, we interpret it as an upper limit on the true electron temperature.
The reason for this systematic overestimation could be an inaccurate sky subtraction in the \sii$\lambda$4069 line or an unresolved blending with the near \oii$\lambda$4069.89 line.

% compute t^2 for all temperatures
Further, we can directly estimate the amount of temperature variations across the plane of the sky by using Eq.~(12) from \citet{Peimbert1967}, originally for a 3D nebula but here for the 2D projection of the sky plane:
\begin{equation}
t_{\text{ps}}^2\ \text{=}\ \frac{\langle (\text{T}_e\,\text{[K]} - \text{T}_0\,\text{[K]})^2 \rangle}{\text{T}_0\,\text{[K]}^2}
\end{equation}
where $t_{\text{ps}}^2$ is a parameter for temperature fluctuations across the plane of the sky, T$_0$ is the mean electron temperature, and T$_e$ is the local electron temperature.
Applying this to the electron temperature gradients given above, we calculate:
$t_{\text{ps}}^2$(\nii)~=~(8.5~$\pm$~1.2)~$\times$~10$^{-5}$,
$t_{\text{ps}}^2$(\oii)~=~(8.6~$\pm$~1.0)~$\times$~10$^{-5}$,
$t_{\text{ps}}^2$(\sii)~=~(4.3~$\pm$~0.6)~$\times$~10$^{-4}$ and
$t_{\text{ps}}^2$(\siii)~=~(2.0~$\pm$~0.2)~$\times$~10$^{-4}$.
These values indicate that the temperature fluctuations are very small, close to zero, consistent with what is seen directly in the temperature maps (see Fig.~\ref{fig:temperature}).

%%%%%%%%%%%%%%%%%%%%%%%%%%%%%%%%%%%%%%%%%%%%%%%%%%%%%%%%%%%%%%%%%%%%%%%%%%%%%%%%%%%%%%%%%%%%%%%%%%%%%%%%%%%%%%%%%%%%%%%

\subsection{Spatially Resolved Oxygen Abundance}
\label{subsec:abundances}

% general structure and ionization zones
Maps of the oxygen abundances and radial oxygen abundance trends are presented in Fig.~\ref{fig:Oabundances}.
It can be seen that the overall abundance of O$^{+}$ is much higher than that of O$^{2+}$, indicating a low level of ionization in the nebula.
Moreover, the abundance of O$^{2+}$ shows a pronounced negative radial gradient with a higher abundance of $\sim$~7.5 in the central regions, where we expect the high ionization zone, and lower values $\sim$~7.0 in the outer regions.
The abundance of O$^{+}$ exhibits the opposite trend, with values around 8.4 in the central region to abundances around 8.6 in the outer region, where we expect the low-ionization zone.
This behavior of the O$^{2+}$ and O$^{+}$ abundances neatly delineates the two ionization zones of the Trifid Nebula: The high-ionization zone up to a distance of around 1.8~pc from the central ionizing source, and the low-ionization zone at distances larger than 1.8~pc (see also Fig.~\ref{fig:Oabundances}).

% homogeneity
Analogous to the electron temperatures discussed in Sect.~\ref{subsec:temperature}, the total oxygen abundance was fitted with a linear function (resulting gradient can be found in App.~\ref{app:gradients}).
When considering both the 3$\sigma$ fit uncertainty and the typical uncertainties of individual spaxels (approximately 0.1~dex, see Fig.~\ref{fig:Oabundance_error}), any positive radial gradient cannot be robustly established. 
Azimuthal variations are also smaller than the measurement uncertainties, further supporting a largely homogeneous abundance distribution.
% slight positive radial trend
The slight apparent trend may reflect variations in the electron temperature used to determine the O$^{+}$ abundance. 
While the electron temperature gradients were found to be relatively flat within the uncertainties, here they could contribute to the inferred metallicities ranging from 8.4 in the inner regions to 8.6 in the outskirts.
Such a difference cannot be explained by dust depletion alone, which at solar metallicity accounts for only $\sim$~0.1~dex \citep{Mesa-Delgado2009}.
Overall, the total oxygen abundance appears approximately flat and homogeneous across the nebula, and any apparent gradient should be interpreted with caution, particularly given the uncertainties in T$_e$(\nii) and T$_e$(\oii).

\subsection{Integrated Physical Properties}
\label{subsec:integrated_res}

Tab.~\ref{tab:integrated_results} presents the results for the physical properties of the integrated spectrum (see Sect.~\ref{subsec:integrated_calc}), alongside the mean values from the resolved measurements.
Both sets are compared directly below.

\begin{table}
    \caption{Results of the integrated measurements together with the mean values of the resolved calculations.}
    \centering
    \begin{tabular}{p{2.2cm} p{1.4cm} p{1.9cm} p{1.5cm}}
    \hline
    \\
    \hfill{} Property & Spectrum & Value & Unit \\
    \hline
    \\
    \hfill{} n$_e$(\oii) & integrated \newline resolved & 95 $\pm$ 16 \newline 85 $\pm$ 30 & cm$^{-3}$ \\
    %85 $\pm$ 31 & cm$^{-3}$ \\
    %\\
    \hfill{} n$_e$(\sii) & integrated \newline resolved & 103 $\pm$ 7 \newline 97 $\pm$ 37 & cm$^{-3}$ \\
    %93 $\pm$ 37 & cm$^{-3}$ \\
    %\\
    \hfill{} T$_e$(\nii) & integrated \newline resolved & 8805 $\pm$ 909 \newline 8541 $\pm$ 382 & K \\
    %\newline 8575 $\pm$ 470 & K \\
    %\\
    \hfill{} T$_e$(\oii) & integrated \newline resolved & 7677 $\pm$ 311 \newline 8180 $\pm$ 324 & K \\
    %\newline 8098 $\pm$ 355 & K \\
    %\\
    \hfill{} T$_e$(\sii) & integrated \newline resolved & 11207 $\pm$ 1420 \newline 10024 $\pm$ 905 & K \\
    %\newline 10057 $\pm$ 885 & K \\
    %\\
    \hfill{} T$_e$(\siii) & integrated \newline resolved & 9357 $\pm$ 544 \newline 9035 $\pm$ 374 & K \\
    %\newline 9134 $\pm$ 446 & K \\
    %\\
    \hfill{} T$_e$(He~\textsc{i}) & integrated & 11417 $\pm$ 3292 & K \\
    %\\
    \hfill{} 12+log(O/H) & integrated \newline resolved & 8.49 $\pm$ 0.14 \newline 8.47 $\pm$ 0.09 &  \\
    %\newline 8.51 $\pm$ 0.14 &  \\
    \hfill{} 12+log(O$^{+}$/H) & integrated \newline resolved & 8.46 $\pm$ 0.15 \newline 8.46 $\pm$ 0.12 &  \\
    %0.15 \newline 8.49 $\pm$ 0.17 &  \\
    \hfill{} 12+log(O$^{2+}$/H) & integrated \newline resolved & 7.31 $\pm$ 0.10 \newline 7.38 $\pm$ 0.27 &  \\
    %0.10 \newline 7.35 $\pm$ 0.28 &  \\
    %\\
    \hfill{} $Q_0$ & integrated & 1.8~$\pm$~0.1 & $\times$ 10$^{49}$~s$^{-1}$ \\ %$\gamma$/s \\
    %\\
    %\hfill{} $f_\text{esc}$ & integrated & 60 $\pm$ 11 & [\%] \\
    \hline
    \\
    \end{tabular}
    \label{tab:integrated_results}
\end{table}

% densities
Looking at the different measurements of the electron density, one can see that both views are consistent, given their uncertainties.

% temperatures
Electron temperatures from resolved and integrated data match within 1$\sigma$ for all diagnostics.
% He
Using the integrated spectrum, we were also able to calculate the helium temperature T$_e$(He~\textsc{i})~=~11417~$\pm$~3292~K.
While the nominal value of this recombination line measurement is somewhat higher than the integrated temperatures of T$_e$(\nii), T$_e$(\oii), and T$_e$(\siii), it is still consistent with those temperatures, given the uncertainties.
The He$\lambda$7281 line is well fitted, and no clear blend is visible, although a minor contribution from a residual sky line cannot be entirely ruled out, which could slightly increase the measured flux and thus the derived temperature.
Moreover, the He\,\textsc{i} emission lines may be affected in cases where the assumption of pure Case~B recombination does not hold, as discussed in \citet{Mendez-Delgado2025}. 
Deviations from Case~B conditions can arise due to the absorption of He\,\textsc{i} photons by neutral hydrogen, or as a result of general leakage of ionizing photons from the nebular region.

% abundance
The integrated and mean value of the resolved oxygen abundance are nearly identical with 8.49~$\pm$~0.14 and  8.47~$\pm$~0.09, respectively.

%%%%%%%%%%%%%%%%%%%%%%%%%%%%%%%%%%%%%%%%%%%%%%%%%%%%%%%%%%%%%%%%%%%%%%%%%%%%%%%%%%%%%%%%%%%%%%%%%%%%%%%%%%%%%%%%%%%%%%%

\section{Discussion}
\label{sec:discussion}

Our IFU mapping reveals the full two-dimensional structure of the nebula.
Below, we discuss the fully recovered density, temperature, abundance and ionizing structures.

%%%%%%%%%%%%%%%%%%%%%%%%%%%%%%%%%%%%%%%%%%%%%%%%%%%%%%%%%%%%%%%%%%%%%%%%%%%%%%%%%%%%%%%%%%%%%%%%%%%%%%%%%%%%%%%%%%%%%%%

\subsection{Density Structure}

Our spatially resolved electron density maps of the Trifid Nebula provide one of the most detailed views to date of its internal density structure, revealing both the expected radial gradient and discrete, compact density enhancements, and, for the first time, allowing a direct comparison between resolved and integrated densities in this region.

% gradient
The measurements of the electron densities, reflecting a line-of-sight average of the single-ionized zone, show a negative radial gradient as well as some discrete regions of enhanced density (see Sect.~\ref{subsec:density}).
The inner double-ionized region remains unprobed as it lacks a direct diagnostic.
A negative radial density gradient was expected and also previously observed for other compact \hii\ regions \citep[e.g.,][]{Binette2002, Osterbrock2006, McLeod2016, Jin2023}.
There, clear negative slopes between $-45$~to~$-850$~cm$^{-3}$/pc were measured in \citet{Jin2023}, while \citet{McLeod2016} reported slopes of $-2000$~to~$-100$~cm$^{-3}$/pc and slopes in \citet{Binette2002} can be as extreme as $-2000$~cm$^{-3}$ per 0.1~pc.
Electron density slopes in our measurements range between $-50$ and $-20$~cm$^{-3}$/pc.
Such declining density profiles are consistent with expanding \hii\ regions, where the ionized gas disperses the remnant molecular material \citep{Geen2015}.

% enhanced regions
We identify two unresolved clumps of enhanced density. 
The higher-density region south-east of HD\,164492A (RA~$\approx$~270.61$^\circ$, DEC~$\approx$~-23.07$^\circ$, Fig.~\ref{fig:density}) is also seen in maps from \citet{Kuhn2022}, where a notably higher dust column density was measured compared to other regions in the nebula.
This enhancement spatially corresponds to the TC\,2 molecular column \citep{Hester2004, Rho2008}, a large collection of molecular gas exposed to ionizing photons. 
The resulting ionization produces free electrons, which we detect as higher electron densities.  
A similar mechanism could explain the high electron density north-west of the ionizing star (RA~$\approx$~270.58$^\circ$, DEC~$\approx$~-23.02$^\circ$, Fig.~\ref{fig:density}), also reported as an obscured high-density region by \citet{Copetti2000} and spatially associated with a dense dust concentration (see Fig.~\ref{fig:rubin}). 
These compact, high-density structures probably represent remnants of the original molecular cloud being eroded by the advancing ionization front.

% compare to literature
Overall, our measurements agree well with the literature's long-slit measurements, but provide a more complete and comprehensive view of the nebula.
Comparing the absolute values of electron density, \citet{Garcia-Rojas2006} measured n$_e$(\oii)~=~240~$\pm$~70~cm$^{-3}$ and n$_e$(\sii)~=~320~$\pm$~130~cm$^{-3}$ close to the ionizing star. 
Moreover, \citet{Copetti2000} measured a median n$_e$(\sii) of 161~cm$^{-3}$ and a maximum of around 330~cm$^{-3}$ using multiple long-slit positions also close to the center of M\,20.
% GR: 17 arcsec north and 10 arcsec east of HD164492 -> around 0.136 pc
% Cop: RA=18 02 23.1   DEC=−23 01 59 and RA=18 02 23.1    DEC=−23 01 34 -> 0.07 pc and 0.13 pc
As both studies used long-slit observations very close to the central ionizing star (less than 0.14~pc in distance, which is 0.10~pc lower than our spatial resolution), their results probe conditions in the immediate stellar environment, whereas our integrated and mean resolved values characterize a much more extended part of the nebula. 
The two approaches, therefore, trace different spatial scales and are not directly comparable.
Instead, we will compare to the electron densities we see closest to the long-slit position, with values of: n$_e$(\oii)~=~216~$\pm$~33~cm$^{-3}$ and n$_e$(\sii)~=~257~$\pm$~25~cm$^{-3}$, which are in the 1$\sigma$ range.
A study by \citet{Rodriguez1999} measured the \sii\ electron density with three long-slit positions in the south-east direction, farther away from the center.
Their first slit position is very close to the TC\,2 molecular cloud and shows an electron density of 360~cm$^{-3}$, where we reach in the corresponding spaxel a similar value of n$_e$(\sii)~=~359~$\pm$~37~cm$^{-3}$.
The second long-slit was placed north-east of TC\,2, measuring 120~cm$^{-3}$. 
There we measure a density of n$_e$(\sii)~=~157~$\pm$~19~cm$^{-3}$, being in the 2$\sigma$ range.
A third slit measured an electron density of 340~cm$^{-3}$ south of the ionizing star and south-west of TC\,2 inside a dusty filament, where we reach a much lower value of n$_e$(\sii)~=~217~$\pm$~22~cm$^{-3}$ outside the 3$\sigma$ range.
However, \citet{Rodriguez1999} did not provide uncertainties to their measured values, limiting the comparison.

% integrated vs resolved
Assessing the impact of inhomogeneities on the integrated measurement, we find no significant discrepancy between our resolved and integrated density measurements in the Trifid Nebula. 
Such a direct comparison is rare and provides a valuable perspective on how local structures influence global nebular diagnostics.  
The observed density distribution of M\,20 is consistent with evolving \hii\ regions in a dense, clumpy molecular environment.
The combination of smooth radial gradients and compact density peaks likely reflects ongoing photo-evaporation and the gradual dissolution of the original molecular and dusty structures.

%%%%%%%%%%%%%%%%%%%%%%%%%%%%%%%%%%%%%%%%%%%%%%%%%%%%%%%%%%%%%%%%%%%%%%%%%%%%%%%%%%%%%%%%%%%%%%%%%%%%%%%%%%%%%%%%%%%%%%%

\subsection{Temperature Structure}

Our spatially resolved electron temperature measurements of M\,20 provide the first detailed two-dimensional view of thermal conditions across the nebula, enabling a direct assessment of temperature homogeneity, a resolved comparison with integrated values, and a new spatial approach to estimating temperature inhomogeneities using $t_{\text{ps}}^2$.

% homogeneous, no gradient
Despite targeting potential temperature inhomogeneities, our measurements of the electron temperature (see Sect.~\ref{subsec:temperature}) reveal remarkably homogeneous conditions without any significant radial gradients.
Compared to the more extended \hii\ regions typically targeted in extragalactic studies, a compact region such as M\,20 would be unresolved and easily blended with diffuse ionized gas in the integrated spectra of external galaxies.
This also suggests that temperature inhomogeneities may not be present, or at least not significant, in all types of \hii\ regions.

Although outwardly increasing temperatures might be expected due to the penetration of high-energy ionizing photons into the outer regions of the nebula \citep{Osterbrock2006}, the thermal balance is governed not only by heating but also by spatially varying cooling processes. 
In particular, changes in the ionization structure can substantially affect the local cooling rates. 
As a result, only detailed photoionization models can determine whether the electron temperature increases or decreases with radius. 
In the case of a model as described in App.~\ref{app:toymodel}, a slight outward increase in electron temperature is predicted.
The lack of a temperature rise at larger distances from the central ionizing star may result from higher uncertainties (>~1000~K, see Fig.~\ref{fig:temperature_error}) in the outermost spaxels.
But, as mentioned above, positive gradients in electron temperature are not crucial for \hii\ regions and homogeneous temperatures are also measured  e.g., four \hii\ regions in the Large and Small Magellanic Clouds \citep{Jin2023}.

% compare to Garcia-Rojas
Now we compare our individual electron temperature measurements to those reported in the literature, particularly focusing on the work of \citet{Garcia-Rojas2006} and \citet{Rodriguez2005}. 
For the \nii\ temperature, \citet{Garcia-Rojas2006} reported T$_e$(\nii)~=~8375~$\pm$~400~K close to the nebular center. 
Our measured value of T$_e$(\nii)~=~9351~$\pm$~597~K is consistent within 2$\sigma$.
Similarly, for \oii, \citet{Garcia-Rojas2006} found T$_e$(\oii)~=~8275~$\pm$~350~K, while we measure  T$_e$(\oii)~=~8144~$\pm$~191~K, which is in agreement within 1$\sigma$. 
For \sii, \citet{Garcia-Rojas2006} reported T$_e$(\sii)~=~6950~$\pm$~350~K, while we measure  T$_e$(\sii)~=~8598~$\pm$~1435~K, which is in agreement within 2$\sigma$. 
In the case of \siii, our measurement of T$_e$(\siii)~=~9190~$\pm$~423~K lies in the 2$\sigma$ range with the value of 8300~$\pm$~400~K reported by \citet{Garcia-Rojas2006}.

% Different observations & methods
These discrepancies may arise from differences in observational data quality or analytical methodology.
While LVM spectra have a resolution of R~$\sim$~4000, the long-slit spectra of UVES, used in \citet{Garcia-Rojas2006}, have R~$\sim$~8800.
Also in their work, \citet{Garcia-Rojas2006} used an $R_V$~=~3.1 and a Seaton reddening law \citep{Seaton1979} while we use $R_V$~=~5.5 and the modified Cardelli law \citep{Cardelli1989, Blagrave2007} as pointed out in Sect.~\ref{subsec:reddening}.
However, testing their reddening setup on our LVM data does not cause a significant change in the measured electron temperatures.
Moreover, an updated set of atomic data was used in our calculations.
The sum of these differences in the analysis could be responsible for major differences between these studies.
% Rodriguez comparison
We also compare our \nii\ temperature measurements to those reported by \citet{Rodriguez1999} at three different positions across the nebula. 
They report T$_e$(\nii)~=~8500~K, 8300~K, and 8600~K, while our corresponding values are 8530~$\pm$~401~K, 8773~$\pm$~659~K, and 8646~$\pm$~404~K, respectively. 
All of our measurements lie within the 1$\sigma$ range of the values by \citet{Rodriguez1999}. 
However, \citet{Rodriguez1999} did not quote uncertainties, limiting a stricter comparison.

% radio temperature
Using radio recombination lines, \citet{Khan2024} measured electron temperatures as a function of the galactocentric radius and calculated a relation. 
Applying the galactocentric distance for M\,20 of 6.79~kpc \citep{Mendez-Delgado2022b}, the expected electron temperature would be 7269~$\pm$~1700~K.
This value is in 1$\sigma$ range with our resolved and integrated measurements of T$_e$(\nii) and T$_e$(\oii), and still in 2$\sigma$ range of our measurements of T$_e$(\sii), T$_e$(\siii) and T$_e$(He~\textsc{i}) (see Sect.~\ref{subsec:temperature} and Tab.~\ref{tab:integrated_results}).

% integrated vs resolved
For all electron temperature measurements, the values of integrated and resolved views are all close and within the 1$\sigma$ range.

% t^2
Lastly, \citet{Garcia-Rojas2006} calculated, using different methods, a $t^2$ value ranging between 0.017~$\pm$~0.010 and 0.049~$\pm$~0.019, while most of our $t_{\text{ps}}^2$ measurements are much smaller and the order of 10$^{-4}$ to 10$^{-5}$. 
While \citet{Garcia-Rojas2006} measured the inhomogeneity $t^2$ using different calculations of the electron temperature from long-slit observations (as these temperatures trace different physical zones within the nebula, see Sect.~\ref{sec:introduction}), we can infer it more reliably by directly using the measured temperature gradients from spatially resolved LVM data across the plane of the sky.

%%%%%%%%%%%%%%%%%%%%%%%%%%%%%%%%%%%%%%%%%%%%%%%%%%%%%%%%%%%%%%%%%%%%%%%%%%%%%%%%%%%%%%%%%%%%%%%%%%%%%%%%%%%%%%%%%%%%%%%

\subsection{The Impact of Inhomogeneities on the Metallicity}

Our spatially resolved abundance analysis enables, for the first time in M\,20, a direct comparison between resolved and integrated metallicity measurements.
This reveals how local inhomogeneities, or their absence, impact abundance determinations and provides new insight into the longstanding debate over gas versus stellar metallicities.

% inhomogeneities
A main goal of this study was to understand the influence of density and temperature inhomogeneities on the measured metallicity in unresolved, integrated \hii\ regions.
Despite showing no relevant temperature inhomogeneities, our resolved measurements of the oxygen abundance seem to follow a slight positive radial gradient.
But due to uncertainties, the radial gradient might be consistent with a flat profile.
% Add referee comment
While our spatially resolved analysis shows that the density variation exceeds the temperature and abundance variations within M\,20, the dominant source of inaccuracy for metallicity calibrations in unresolved extragalactic regions is likely the variation in ionization parameter and density structure \citep{Kewley2019a, Mendez-Delgado2023a}. 
The integrated spectra mix emission from high- and low-ionization zones as well as DIG, shifting strong-line ratios away from single-zone calibration relations. 
However, abundances measured with auroral line electron temperatures can also be affected by temperature inhomogeneities, but for typical extragalactic strong-line methods the ionization and density structure as well as DIG contamination drive the main systematic uncertainties.

% compare to Garcia-Rojas
Compared to other studies, our resolved and integrated measurements of the absolute oxygen abundance (8.47~$\pm$~0.09 and 8.49~$\pm$~0.14 respectively) are lower than the calculation of \citet{Garcia-Rojas2006} which includes temperature inhomogeneities, resulting in a total oxygen abundance of 8.67~$\pm$~0.03 (we only take into account the O$^{+}$ and O$^{2+}$ abundance measurements to keep it most similar to our study).
However, our measured abundances match well with the homogeneous approach by \citet{Garcia-Rojas2006}, resulting in a total oxygen abundance of 8.53~$\pm$~0.03.
% direct spaxel comparisons
When we compare the total oxygen abundance in the spaxel corresponding to the long-slit position of \citet{Garcia-Rojas2006}, our measurement is 8.39~$\pm$~0.08.
This value falls within the 2$\sigma$ range of their result using the homogeneous approach, but lies outside the 3$\sigma$ range when compared with their inhomogeneous assumption.
% compare to Rodriguez
We can also compare our results to the different long-slit positions of \citet{Rodriguez1999}.
In the first slit position, the combined O$^{+}$ and O$^{2+}$ abundance is 8.22, which is outside the 3$\sigma$ range of our spaxel value of 8.44~$\pm$~0.05.
The third slit position shows a similar oxygen abundance of 8.20, while in the corresponding spaxel, we measure again 8.44~$\pm$~0.05, which is also outside the 3$\sigma$ range.
Only in the second slit position with a value of 8.32, we are measuring 8.38~$\pm$~0.08 and are within the 1$\sigma$ range.
However, similar to Sect.~\ref{subsec:density} and \ref{subsec:temperature}, \citet{Rodriguez1999} did not give uncertainties for their measurements, which limits a qualitative comparison.

\begin{figure}
    \centering
    \includegraphics[width=0.9\linewidth]{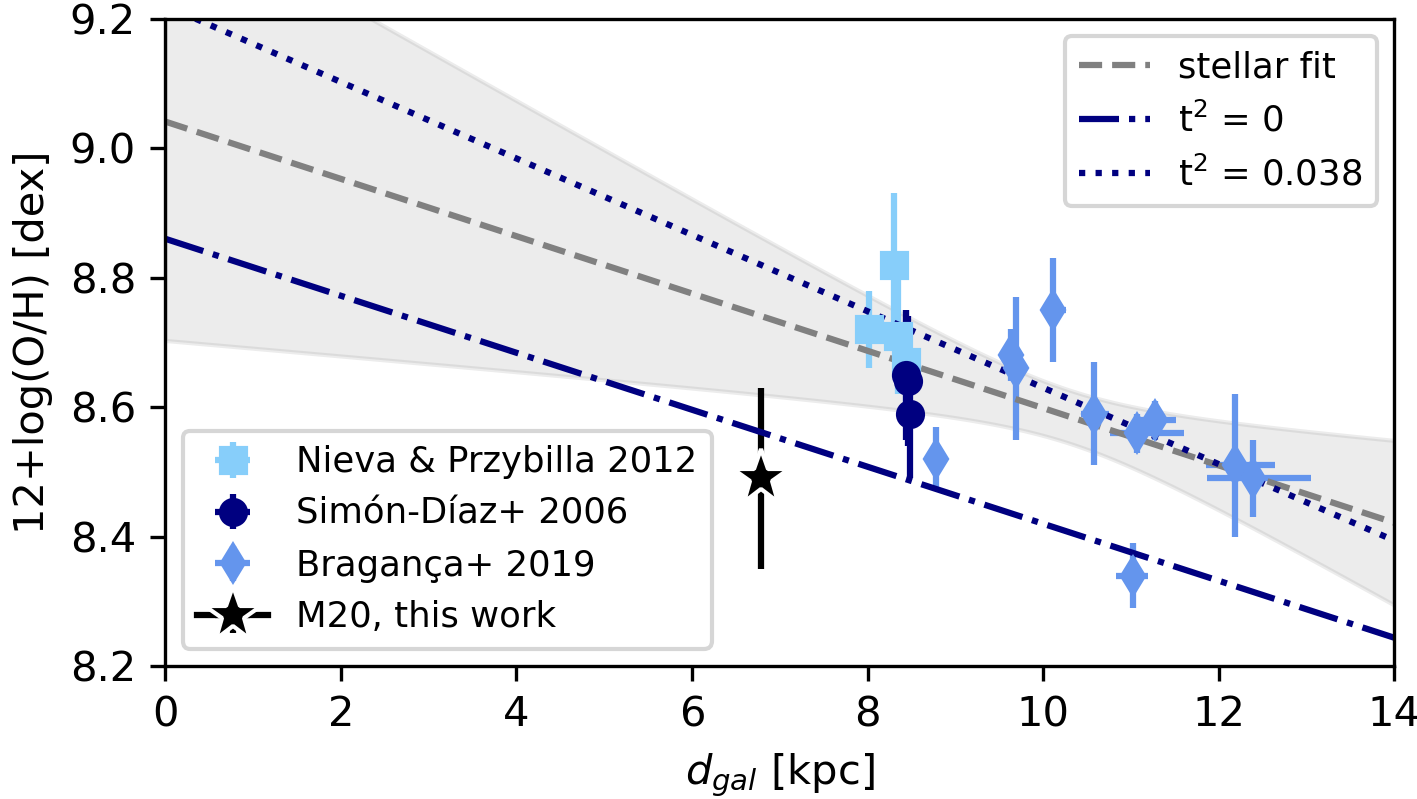}
    \caption{Oxygen abundance trend calculated from B\,V-stellar measurements by \citet{Simon-Diaz2006, Nieva2012, Braganca2019}.
    The fit is illustrated by a grey dashed line, whereas the shaded region indicates the 3$\sigma$ uncertainty of the fit.
    Fits from \citet{Mendez-Delgado2022b} for t$^2$~=~0 and t$^2$~=~0.038 are shown as dark blue dash-dotted and dotted lines, respectively.
    The position of the nebular oxygen abundance of M\,20 calculated in this work (8.49~$\pm$~0.14, from the integrated view) is indicated by the black star.}
    \label{fig:stellarabundances}
\end{figure}

% compare to stellar abundances
Comparing our measured mean resolved and integrated gaseous oxygen abundances to the stellar one of HD\,164492A (8.56~$^{+0.3}_{-0.12}$, \citealt{Martins2015}), we find a good match between both abundance measurements.
% fit to BV stars
However, as the abundance measurement of O-stars is very difficult and leads to large uncertainties, we also present in Fig.~\ref{fig:stellarabundances} a relationship between B-dwarf stellar abundances and the galactocentric radii.
We used this method as a second approach, besides the directly measured oxygen abundance of HD\,16642A by \citet{Martins2015}, to gain another estimate of the expected stellar oxygen abundance in the Trifid Nebula.
Therefore, we used measured oxygen abundances of B\,V stars by \citet{Simon-Diaz2006, Nieva2012, Braganca2019} and distances for these objects as measured by \citet{Bailer-Jones2021}.
We fitted a linear relation to the distribution in an abundance-distance plane and derived the following relationship:
\begin{align}
12 + \log(\text{O/H}) &= (-0.04 \pm 0.01)\,[\text{dex/pc}]\,(d_{\text{gal}} \pm 1.46)\,[\text{kpc]} \nonumber\\
                      &\quad + (9.04 \pm 0.11)
\label{eq:stellarab}
\end{align}
where $d_{\text{gal}}$ is the galactocentric distance of the object.
Using Eq.~\eqref{eq:stellarab}, we derive for M\,20 at a galactocentric distance of 6.79~$^{+0.09}_{-0.08}$~kpc \citep{Mendez-Delgado2022b} an oxygen abundance of 8.74~$\pm$~0.04.
It can be seen that the estimated abundance from B-dwarfs is over 0.2~dex higher than the one measured for HD\,164492A and M\,20. 
In principle, the oxygen abundance in massive stars (O and early B types) could be lower due to CNO-equilibrium processing.
Also, we lack abundance measurements of B\,V stars at distances below 7~kpc, making the prediction much harder.
If the stellar oxygen abundance was indeed higher than the measured gas abundance, this might suggest oxygen depletion into dust, consistent with the prominent dust lanes observed across M\,20 (e.g., Fig.~\ref{fig:rubin}). 
Nevertheless, a discrepancy of 0.2~dex exceeds the expected contribution from dust depletion, which is limited to around 0.1~dex \citep{Mesa-Delgado2009, Simon-Diaz2011_orion}. 
For reference, the solar oxygen abundance is 8.69~$\pm$~0.04 \citep{Asplund2021}, and it would be reasonable to assume that M\,20 appears slightly more metal-rich than the Sun, given its location closer to the Galactic center.

% compare to eduardos fits 2022b
When comparing our results to the oxygen abundance gradients obtained through multiple Galactic \hii\ regions reported by \citet{Mendez-Delgado2022b} (their Tab.~2 and Fig.~4), we find good agreement with the gradient obtained assuming $t^2 = 0$. 
Although our measurement of M\,20's oxygen abundance lies outside the 1$\sigma$ range of the fit assuming $t^2 = 0.038$, this latter fit shows better agreement with the gradient derived from stellar abundances.

% general implications
Interestingly, both the gas-phase and O-star abundances in M\,20 are relatively low compared with the Galactic gradient traced by B-dwarfs and multiple \hii\ regions \citep[e.g.,][]{Mendez-Delgado2022b}, as illustrated in Fig.~\ref{fig:stellarabundances}.  
This raises the questions of whether M\,20 reflects a local region of lower metallicity, potentially requiring an influx of relatively pristine gas or retaining signatures of its initial, less enriched environment.  
And if in such compact, relatively quiescent \hii\ regions, one should not expect significant temperature inhomogeneities, which would be consistent with the remarkably homogeneous electron temperature we measure across M\,20.

%%%%%%%%%%%%%%%%%%%%%%%%%%%%%%%%%%%%%%%%%%%%%%%%%%%%%%%%%%%%%%%%%%%%%%%%%%%%%%%%%%%%%%%%%%%%%%%%%%%%%%%%%%%%%%%%%%%%%%%

\subsection{Exploring the Ionization Structure of M\,20}
\label{subsec:discussIonization}

Different approaches to estimate the amount of ionizing photons emitted by the central stellar source (see Sect.~\ref{sec:Q0}) yield different values.
% discrepancy in Q_0
The measured value (using the tailored model) of $Q_0~=~(1.8~\pm~0.1)~\times~10^{49}$~s$^{-1}$ is considerably higher than $Q_0$~=~4.07~$\times$~10$^{48}$~s$^{-1}$ and $Q_0$~=~7.6~$\times$~10$^{48}$~s$^{-1}$, the theoretical ionizing rates of an O\,7.5\,V and an O\,6.5\,V star, respectively \citep{Martins2005}.
Obtaining a lower $Q_0$ would require physical conditions that are considered unlikely for this system.
However, our measured value matches well with the ionizing photon rate previously reported by \citet{Binder2018} of $Q_0$~=~(1.5~$\pm$~0.4)~$\times$~10$^{49}$~s$^{-1}$.
% Discrepancy with theory
These discrepancies from the theoretical ionizing rates may be attributed to several factors:

% mis-classification?
The most straightforward interpretation of this discrepancy is that the ionizing source, HD\,164492A, is likely of an earlier spectral type than the recent O\,7.5\,V classification suggests.  
The measured effective temperature by \citet{Martins2015} of 38000~K leads to the assumption of an O\,6.5\,V star, while from our measurements, the ionizing flux is consistent with an O\,5\,V star \citep{Martins2005}.  
HD\,164492 is also a complex multiple system including O\,7\,V, B\,6\,V, A\,2\,Ia, and possibly a Be star \citep{Rho2004, Rho2006, Binder2018}, but these additional stars contribute negligibly to the total ionizing flux, as the earliest spectral type dominates by orders of magnitude \citep[e.g.,][]{Ramachandran2019}.  
% binarity?
Previous suggestions of binarity for HD\,164492A \citep{Stickland2001} could affect the inferred spectral type, but searches for a companion remain inconclusive.

% dust
Moreover, prominent dust lanes are observed across the Trifid Nebula (Fig.~\ref{fig:rubin}), so inferring $Q_0$ from H$\alpha$ under the assumption of a dust-free nebula would underestimate the true ionizing photon output (see App.~\ref{app:toymodel}).  
Our \textsc{cloudy} toy model suggests that the difference between the classical Strömgren radius and the observed nebular size can be explained by the combined effects of dust opacity and density inhomogeneities.
When compared to \citet{Kennicutt1998}, the higher ratio of $Q_0$/$L$(H$\alpha$)~=~2.1~$\times$~10$^{12}$~erg$^{-1}$ obtained by the model further highlights the influence of dust, while regions with lower gas coverage may allow some ionizing photons to escape, so the nebula may not be fully radiation-bounded in all directions.

%%%%%%%%%%%%%%%%%%%%%%%%%%%%%%%%%%%%%%%%%%%%%%%%%%%%%%%%%%%%%%%%%%%%%%%%%%%%%%%%%%%%%%%%%%%%%%%%%%%%%%%%%%%%%%%%%%%%%%%

\section{Conclusions}
\label{sec:conclusion}

The Trifid Nebula (M\,20), a compact, nearly spherical \hii\ region, offers an ideal laboratory to study the internal physical structure of a simple Strömgren sphere ionized by the mid-type O-dwarf HD\,164492A.
Its simplicity in ionizing structure, along with the presence of current star formation, makes M\,20 a benchmark system for testing ionization and abundance diagnostics.

% Observations
We observed M\,20 using the LVM as part of SDSS-V \citep{Perruchot2018, Kollmeier2019, Drory2024, Blanc2024, Herbst2024, Kollmeier2025}, which provides wide-field integral field spectroscopy with a resolution of R~$\sim$~4000 across 3600–9800\,\AA.
With a spatial resolution of 0.24\,pc, we spatially resolved the ionized gas and measured physical properties across the nebula. 

% density
Our spatially resolved measurements of the electron density, temperature, and oxygen abundance reveal a detailed internal view of the nebula. 
We identify a clear negative radial electron density gradient and two distinct clumps of enhanced density, each spatially coincident with dusty or molecular features (see Sect~\ref{subsec:density}). 
These structures likely result from interactions between ionizing photons and dense gas, highlighting the interplay between local environmental conditions and nebular structure.

% temperatures
In contrast, electron temperatures seem remarkably uniform across the nebula (see Sect~\ref{subsec:temperature}). 
This absence of a clear positive temperature gradient, which is often (but not always) expected from ionization theory, may be attributed in part to uncertainties in the outermost regions but also depends on factors such as metallicity and the shape of the ionizing spectral energy distribution. 
However, it is consistent with findings in other homogeneous \hii\ regions \citep[e.g.,][]{Jin2023}. 
Our spatially resolved approach allows us to estimate temperature inhomogeneities using the $t^2$ formalism \citep{Peimbert1967} directly from the measured gradients, revealing significantly lower values than those reported in long-slit studies. 

% abundances
When examining oxygen abundances, we find that resolved and integrated measurements agree closely, with weak evidence for a positive radial metallicity gradient (see Sect.~\ref{subsec:abundances}). 
Our measurements are consistent with other homogeneous analyses in the literature \citep{Garcia-Rojas2006, Mendez-Delgado2022b} and lower than inhomogeneity-corrected estimates. 
Furthermore, we compare our nebular abundance to the stellar oxygen abundance of HD\,164492A, finding general agreement. 
However, when comparing to an extrapolated oxygen abundance trend based on B-dwarfs, our derived oxygen abundance is about 0.2~dex lower than expected for the galactocentric distance of M\,20.

% ionizing structure
The ionization structure of M\,20 appears to exhibit a higher ionizing photon rate than would be expected from a single O\,7.5\,V star. 
This discrepancy likely reflects a combination of factors, including a possible misclassification of the central source HD\,164492A, which may be an earlier-type or binary system, or other modeling assumptions about the physical conditions within the nebula. 
Our \textsc{cloudy} toy model further suggests that dust opacity and density inhomogeneities play a significant role in shaping the observed nebular size and ionization balance, indicating that M\,20 is not fully radiation bounded in all directions.

Overall, this study demonstrates the power of spatially resolved spectroscopy in disentangling the physical and chemical structure of Galactic \hii\ regions. 
Our results show that integrated measurements can reliably trace average nebular conditions. 
Yet, even in cases of simple geometry, resolved maps reveal complex small-scale features and offer more direct insight into the role of local inhomogeneities. 
With the capabilities of LVM, we establish a framework for interpreting unresolved spectra of distant galaxies and for refining abundance diagnostics through detailed benchmarking in nearby star-forming regions like the Trifid Nebula.

%%%%%%%%%%%%%%%%%%%%%%%%%%%%%%%%%%%%%%%%%%%%%%%%%%%%%%%%%%%%%%%%%%%%%%%%%%%%%%%%%%%%%%%%%%%%%%%%%%%%%%%%%%%%%%%%%%%%%%%

\begin{acknowledgements}
      % Referee
      The authors thank the anonymous referee for the quick and helpful review.
      % Authors
      KK, JEMD, EE, and NS acknowledge funding from the Deutsche Forschungsgemeinschaft (DFG, German Research Foundation) in the form of an Emmy Noether Research Group (grant number KR4598/2-1, PI Kreckel) and the European Research Council’s starting grant ERC StG-101077573 (“ISM-METALS"). 
      JEMD, CM and CRZ thank the support by SECIHTI CBF-2025-I-2048 project ``Resolviendo la Física Interna de las Galaxias: De las Escalas Locales a la Estructura Global con el SDSS-V Local Volume Mapper''. 
      CM acknowledges the support of the grant UNAM/DGAPA/PAPIIT IG101223.
      C R.-Z. acknowledges support from grant UNAM-DGAPA-PAPIIT IG101723.
      E. J. acknowledges support from the interdisciplinary project Millennium Nucleus for the Evolutionary Reconstruction of the InterStellar medium (ERIS NCN2021\_017). 
      OE acknowledges funding from the Deutsche Forschungsgemeinschaft (DFG, German Research Foundation) -- project-ID 541068876.
      AACS is supported by the Deutsche Forschungsgemeinschaft (DFG, German Research Foundation) in the form of an Emmy Noether Research Group -- Project-ID 445674056 (SA4064/1-1, PI Sander). AACS further acknowledges funding by the Federal Ministry for Economic Affairs and Climate Action (BMWK) via the German Aerospace Center (Deutsches Zentrum f\"ur Luft- und Raumfahrt, DLR) grant 50 OR 2503 (PI: Sander).
      SFS thanks the support by UNAM PASPA – DGAPA and the SECIHTI CBF-2025-I-236 project.
      J.G.F-T gratefully acknowledges the grants support provided by ANID Fondecyt Postdoc No. 3230001 (Sponsoring researcher), and from the Joint Committee ESO-Government of Chile under the agreement 2023 ORP 062/2023.
      G.A.B.acknowledges the support from the ANID Basal project FB210003.

      % standard SDSS-V
      Funding for the Sloan Digital Sky Survey V has been provided by the Alfred P. Sloan Foundation, the Heising-Simons Foundation, the National Science Foundation, and the Participating Institutions. SDSS acknowledges support and resources from the Center for High-Performance Computing at the University of Utah. SDSS telescopes are located at Apache Point Observatory, funded by the Astrophysical Research Consortium and operated by New Mexico State University, and at Las Campanas Observatory, operated by the Carnegie Institution for Science. The SDSS website is \url{www.sdss.org}.
      SDSS is managed by the Astrophysical Research Consortium for the Participating Institutions of the SDSS Collaboration, including Caltech, The Carnegie Institution for Science, Chilean National Time Allocation Committee (CNTAC) ratified researchers, The Flatiron Institute, the Gotham Participation Group, Harvard University, Heidelberg University, The Johns Hopkins University, L'Ecole polytechnique f\'{e}d\'{e}rale de Lausanne (EPFL), Leibniz-Institut f\"{u}r Astrophysik Potsdam (AIP), Max-Planck-Institut f\"{u}r Astronomie (MPIA Heidelberg), Max-Planck-Institut f\"{u}r Extraterrestrische Physik (MPE), Nanjing University, National Astronomical Observatories of China (NAOC), New Mexico State University, The Ohio State University, Pennsylvania State University, Smithsonian Astrophysical Observatory, Space Telescope Science Institute (STScI), the Stellar Astrophysics Participation Group, Universidad Nacional Aut\'{o}noma de M\'{e}xico, University of Arizona, University of Colorado Boulder, University of Illinois at Urbana-Champaign, University of Toronto, University of Utah, University of Virginia, Yale University, and Yunnan University.
\end{acknowledgements}

\bibliographystyle{aa_url} 
\bibliography{bib.bib} 

\begin{appendix}
\appendix
\counterwithin{figure}{section}

\section{Integrated Emission Line Fluxes}
\label{app:fluxes}

In Tab.~\ref{tab:fluxes}, we give all the observed $I(\lambda)$ and reddening corrected $F(\lambda)$ emission line fluxes for the fully integrated spectrum of M\,20 as calculated in Sect.~\ref{subsec:integrated_calc}, together with the uncertainties.
All fluxes are normalized to the H$\beta$~=~100 with $I($H$\beta)$~=~1.60~$\times$~10$^{-9}$~erg/s/cm$^2$ and $F($H$\beta)$~=~1.16~$\times$~10$^{-8}$~erg/s/cm$^2$.

\begin{table}[h!]
    \caption{Observed $I(\lambda)$ and reddening corrected $F(\lambda)$ fluxes for various emission lines at weavelengths $\lambda$ used in this work of the fully integrated spectrum. 
    Uncertainties for each F($\lambda$) are given in percent.}
    \centering
    \begin{tabular}{ccccc}
    \hline
    \\
    \hfill{} Line [$\si{\angstrom}$] & Ion & $I(\lambda)$ & $F(\lambda)$ & Error [\%] \\
    \hline
    \\
    \hfill{} 3727 & \oii & 96.012 & 135.443 & 1.2 \\
    \hfill{} 3729 & \oii & 129.775 & 183.008 & 0.9 \\
    \hfill{} 3771 & H\,\textsc{i} & 1.599 & 2.236 & 37.2 \\
    %\hfill{} 3797 & H\,\textsc{i} & 2.216 & 3.084 & 19.0 \\
    \hfill{} 3835 & H\,\textsc{i} & 3.21 & 4.43 & 15.1 \\
    %\hfill{} 3889 & H\,\textsc{i} & 12.897 & 17.568 & 4.5 \\
    %\hfill{} 3970 & H\,\textsc{i} & 11.392 & 15.188 & 4.3 \\
    \hfill{} 4069 & \sii & 1.964 & 2.544 & 16.7 \\
    \hfill{} 4076 & \sii & 0.439 & 0.567 & 55.5 \\
    \hfill{} 4102 & H\,\textsc{i} & 19.694 & 25.249 & 1.6 \\
    \hfill{} 4340 & H\,\textsc{i} & 38.871 & 46.11 & 1.0 \\
    %\hfill{} 4471 & He\,\textsc{i} & 2.676 & 3.037 & 5.1 \\
    \hfill{} 4861 & H\,\textsc{i} & 100.0 & 100.0 & 0.1 \\
    %\hfill{} 4959 & \oiii & 14.776 & 14.34 & 0.8 \\
    \hfill{} 5007 & \oiii & 47.456 & 45.401 & 0.5 \\
    \hfill{} 5755 & \nii & 1.272 & 0.996 & 22.4 \\
    %\hfill{} 5876 & He\,\textsc{i} & 11.021 & 8.375 & 4.5 \\
    \hfill{} 6312 & \siii & 1.294 & 0.887 & 14.1 \\
    %\hfill{} 6548 & \nii & 45.8 & 29.715 & 38.5 \\
    \hfill{} 6563 & H\,\textsc{i} & 397.338 & 256.899 & 5.7 \\
    \hfill{} 6584 & \nii & 145.697 & 93.746 & 16.3 \\
    \hfill{} 6678 & He\,\textsc{i} & 3.892 & 2.451 & 0.8 \\
    \hfill{} 6717 & \sii & 49.362 & 30.802 & 0.3 \\
    \hfill{} 6731 & \sii & 36.813 & 22.897 & 0.3 \\
    \hfill{} 7281 & He\,\textsc{i} & 0.833 & 0.457 & 26.6 \\
    \hfill{} 7320 & \oii & 3.203 & 1.741 & 9.8 \\
    \hfill{} 7330 & \oii & 2.658 & 1.442 & 12.0 \\
    %\hfill{} 8750 & H\,\textsc{i} & 1.416 & 0.568 & 13.0 \\
    %\hfill{} 8863 & H\,\textsc{i} & 2.902 & 1.14 & 7.7 \\
    %\hfill{} 9015 & H\,\textsc{i} & 2.736 & 1.045 & 7.1 \\
    %\hfill{} 9069 & \siii & 49.925 & 18.877 & 0.5 \\
    \hfill{} 9229 & H\,\textsc{i} & 5.328 & 1.958 & 2.3 \\
    \hfill{} 9531 & \siii & 109.6 & 38.347 & 1.2 \\
    \\
    \hline
    \\
    %\tablenotes{(1) Emission line wavelength in $\si{\angstrom}$; (2) Corresponding ion; (3) Observed fluxes I($\lambda$); (4) Reddening corrected fluxes F($\lambda$); (5) Uncertainties of the reddening corrected fluxes.}
    \end{tabular}
    \label{tab:fluxes}
\end{table}

\section{Modeling Details for the Cloudy Toy Model of M\,20}
\label{app:toymodel}

The photoionization model for M\,20 is obtained using the latest version of \textsc{cloudy} \citep[v25.00, see][]{2025Gunasekera_arXi}. 

% first model
A first sanity-check model is run without dust or filling factor, with a Black Body as ionizing source, at T$_{\text{eff}}$~=~37500~K, with $Q_0=10^{49}$ s$^{-1}$, and a constant hydrogen density set to 100~cm$^{-3}$ and an inner radius of 3~$\times$~10$^{17}$~cm (which leads to a rather filled sphere geometry). 
The model stops at a radius of 2.75~pc, the value obtained from the classical Strömgren radius definition from Eq.~\eqref{eq:Q0}, using n$_e$~=~106~cm$^{-3}$ and T$_e$~=~7115~K.
The corresponding value for $L$(H$\alpha$) is 1.26~$\times$~10$^{37}$~erg/s, and Q$_0$/$L$(H$\alpha$)~=~7.9~$\times$~10$^{11}$~erg$^{-1}$, close to the value of Q$_0$/$L$(H$\alpha$)~=~$7.32~\times~10^{11}$~erg$^{-1}$ derived in Eq.~\eqref{eq:Q0_2}, the departure is mainly due to the way \textsc{cloudy} computes $L$(H$\alpha$).

When predicting the outer radius of a more complex nebula, several additional effects must be considered: 
dust can compete with neutral hydrogen (H$^0$) in absorbing ionizing photons, which reduces the size of the ionized region. 
Conversely, a filling factor smaller than 1.0 tends to increase the apparent size of the ionized region. 
Furthermore, the ratio Q$_0$/$L$(H$\alpha$) is also affected by the presence of dust.

% more complex model
A more complex model is therefore run, where the ionizing spectrum is taken from a \textsc{fastwind} model \citep{1997Santolaya-Rey_aap323, 2005Puls_aap435} extracted from the pyStarburst99 library \citep{2025Hawcroft_arXi}. 
An O-star at solar metallicity with T$_{\text{eff}}$~=~37500~K is chosen. 
Its luminosity is defined by setting $Q_0$~=~1.8~$\times$~10$^{49}$~s$^{-1}$ to reproduce the observed flux of H$\alpha$ (see Sect.~\ref{sec:Q0}). 
The hydrogen density is set to decrease with the radius from 200 to 50 cm$^{-3}$. 
An inner radius of 3~$\times$~10$^{17}$~cm is set, to have the filled sphere expected from the spatial profile of the emission line shown in Fig.~\ref{fig:Halpha}. 
A filling factor of 0.73 is used to let the outer radius reach a value of 3.06~pc, very close to the observed value (see Sect.~\ref{sec:Q0}). 
The abundance set is taken from \citet{Nicholls2017} based on log(O/H)~=~-3.36, with depletion. 
Grains are taken into account using the \textsc{cloudy} \textit{grains ism} command. 
This model aims to reproduce the principal characteristics of the nebula, mainly its global emission of H$\alpha$ and its size and temperature, but not trying to fine-tune the parameters to reproduce all the observables exactly. 
The model predicts $L$(H$\alpha$)~=~8.7~$\times$~10$^{36}$ erg/s, close to the observed value (see Sect.~\ref{sec:Q0}). 
The mean temperature of H$^{+}$ is 8355~K, very consistent with the electron temperature reported in Tab.~\ref{tab:integrated_results}. 
The temperature is actually almost constant within the nebula, with an increase (800~K) in the very inner and outer regions. 

The ratio between the rate of ionizing photons and the luminosity of the nebula is  Q$_0$/$L$(H$\alpha$)~=~2.1~$\times$~10$^{12}$ erg$^{-1}$.
This value is greater than Q$_0$/$L$(H$\alpha$)~=~$7.32~\times~10^{11}$ erg$^{-1}$ obtained from Eq.~\eqref{eq:Q0_2}, mainly due to the presence of dust in our model. 
The actual ratio may be slightly higher if the nebula is partially matter-bounded in some directions.
This results from the [O \,\textsc{i}]$\lambda$6300 emission being unconstrained in our data, but the model predicts an intensity of 0.024 H$\beta$.  
However, the strong observed [S \,\textsc{ii}] emission indicates that most regions of the nebula are close to the recombination front, implying that it is largely ionization-bounded overall.

A dust-free model is also computed with the same density law, chemical composition, and ionizing spectral energy distribution (SED). 
To fit the values of $L$(H$\alpha$) and the outer radius R, the stellar luminosity and filling factor must be tuned to $Q_0~=~6.9~\times~10^{48}$~s$^{-1}$ and $f_{\text{fill}}~=~0.7$. 
This illustrates the importance of dust in the absorption of ionizing SED (here, 60\% of the ionizing SED is absorbed by dust) and in the relationship between $L$(H$\alpha$) and $Q_0$.

% Add referee comment
Our Cloudy toy model was designed to reproduce the observables directly relevant for estimating the ionizing photon rate Q$_0$, namely the H$\alpha$ luminosity, the electron temperature, and the Strömgren radius. 
Although metallicity and ionic abundances do influence the detailed excitation structure of the nebula, reasonable variations in chemical composition do not significantly change these predictions. 
The model does not aim to reproduce all line ratios because doing so would require a fully constrained ionizing SED, a detailed density structure, and assumptions about the dust content, which go beyond the scope of this work. 
However, the simplifications of our toy model do not affect our determination of Q$_0$ or the conclusions drawn from it.

%%%%%%%%%%%%%%%%%%%%%%%%%%%%%%%%%%%%%%%%%%%%%%%%%%%%%%%%%%%%%%%%%%%%%%%%%%%%%%%%%%%%%%%%%%%%%%%%%%%%%%%%%%%%%%%%%%%%%%%

\section{Fitted Temperature and Abundance Gradients}
\label{app:gradients}

The temperature gradients obtained from the linear fits (see Sect.~\ref{subsec:temperature}) are:
\begin{align}
\text{T}_e(\text{\oii})\,[\text{K}]   &= (-77 \pm 49)\,[\text{K/pc}]\,(d \pm 0.7)\,[\text{pc]}  \nonumber\\
                                       &\quad + (8212 \pm 63)\,[\text{K}] \\[3pt]
\text{T}_e(\text{\nii})\,[\text{K}]   &= (-79 \pm 82)\,[\text{K/pc}]\,(d \pm 0.7)\,[\text{pc]}  \nonumber\\
                                       &\quad + (8666 \pm 104)\,[\text{K}] \\[3pt]
\text{T}_e(\text{\siii})\,[\text{K}]  &= (168 \pm 68)\,[\text{K/pc}]\,(d \pm 0.7)\,[\text{pc]}  \nonumber\\
                                       &\quad + (8884 \pm 77)\,[\text{K}] \\[3pt]
\text{T}_e(\text{\sii})\,[\text{K}]   &= (11 \pm 261)\,[\text{K/pc}]\,(d \pm 0.7)\,[\text{pc]}  \nonumber\\
                                       &\quad + (10156 \pm 326)\,[\text{K}]
\label{eq:temperatures}
\end{align}
where $d$ is the radial distance to the central ionizing star HD\,164492A.
The corresponding oxygen abundance gradient derived from a linear fit (see Sect.~\ref{subsec:abundances}) is:
\begin{align}
12 + \log(\text{O/H}) &= (0.07 \pm 0.01)\,[\text{dex/pc}]\,(d \pm 0.7)\,[\text{pc]} \nonumber\\
                      &\quad + (8.35 \pm 0.01)
\label{eq:abundance}
\end{align}

%%%%%%%%%%%%%%%%%%%%%%%%%%%%%%%%%%%%%%%%%%%%%%%%%%%%%%%%%%%%%%%%%%%%%%%%%%%%%%%%%%%%%%%%%%%%%%%%%%%%%%%%%%%%%%%%%%%%%%%

\section{Uncertainties of Resolved Physical Properties}
\label{app:uncertainties}

\begin{figure}[h!]
    \centering
    \includegraphics[width=\linewidth]{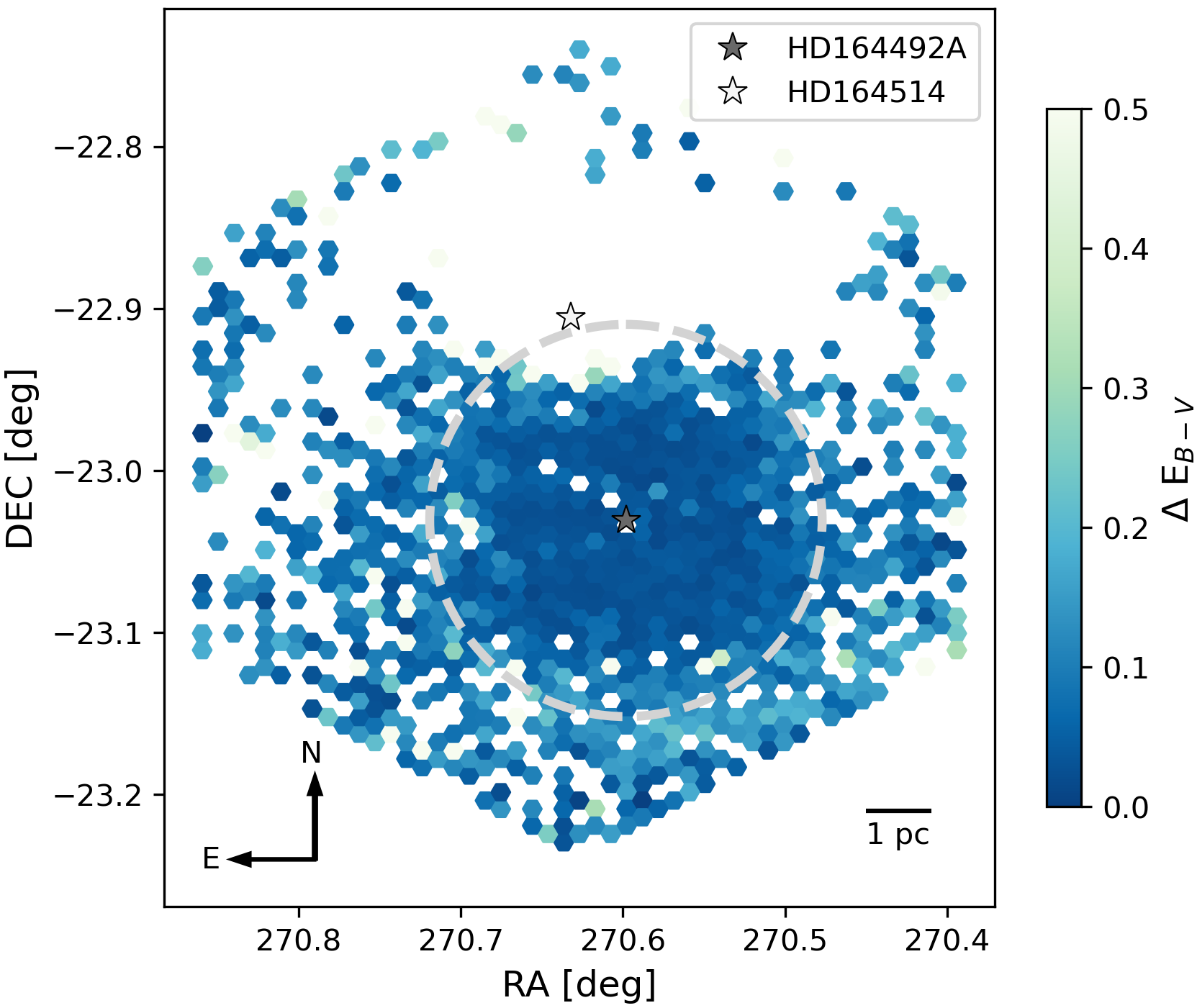}
    \caption{Uncertainties of the dust extinction map shown in Fig.~\ref{fig:reddening}. The position of HD\,164492A is marked with a grey star.
    The extent of the measured Strömgren sphere (Sect.~\ref{sec:Q0}) is shown by a grey dashed circle.}
    \label{fig:reddening_error}
\end{figure}

\begin{figure*}
    \centering
    \includegraphics[width=0.85\linewidth]{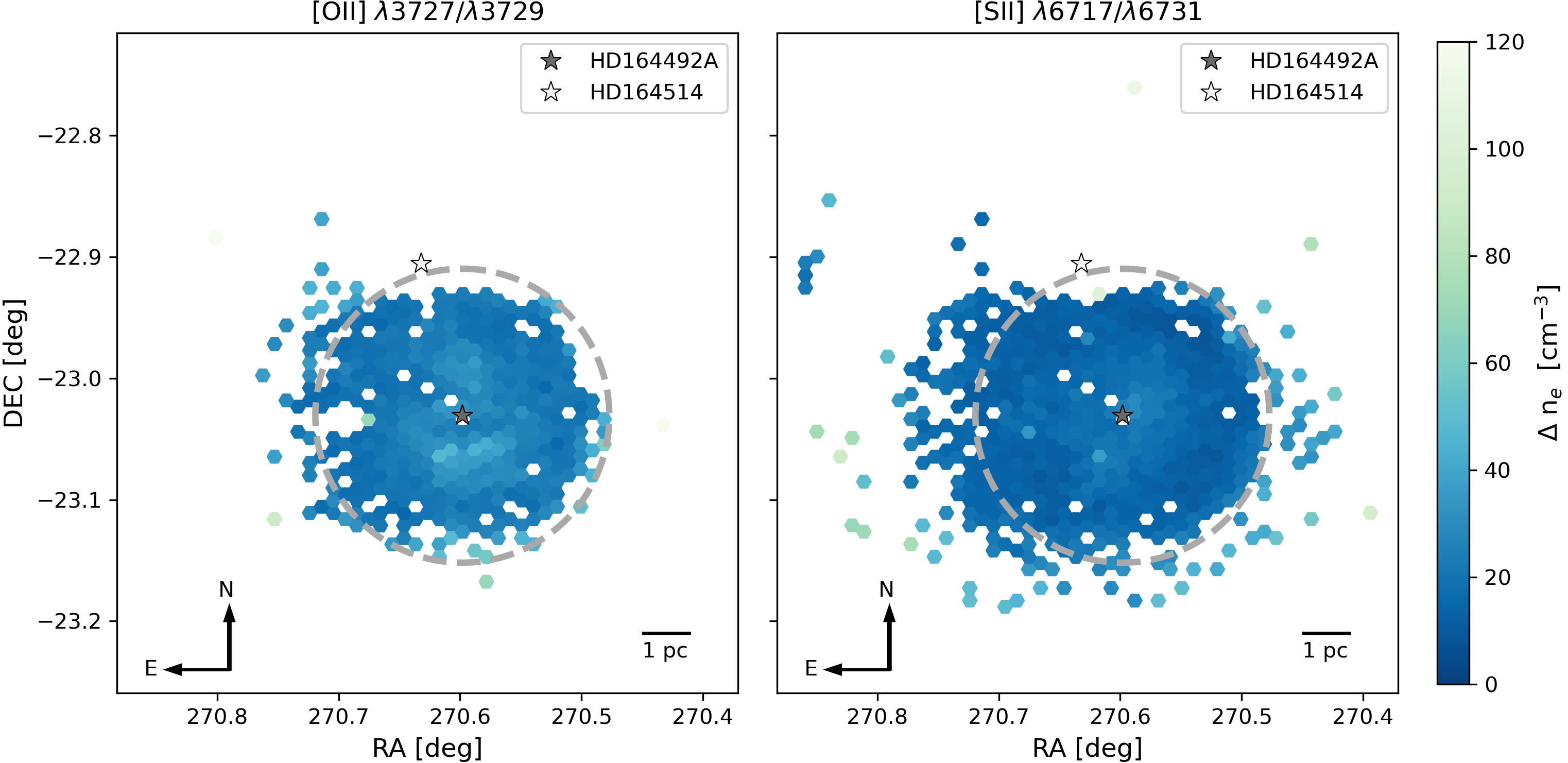}
    \caption{Uncertainties of the electron density maps shown in Fig.~\ref{fig:density}. The position of HD\,164492A is marked with a grey star.
    The extent of the measured Strömgren sphere (Sect.~\ref{sec:Q0}) is shown by a grey dashed circle.}
    \label{fig:density_error}
\end{figure*}

\begin{figure*}
    \centering
    \includegraphics[width=1\linewidth]{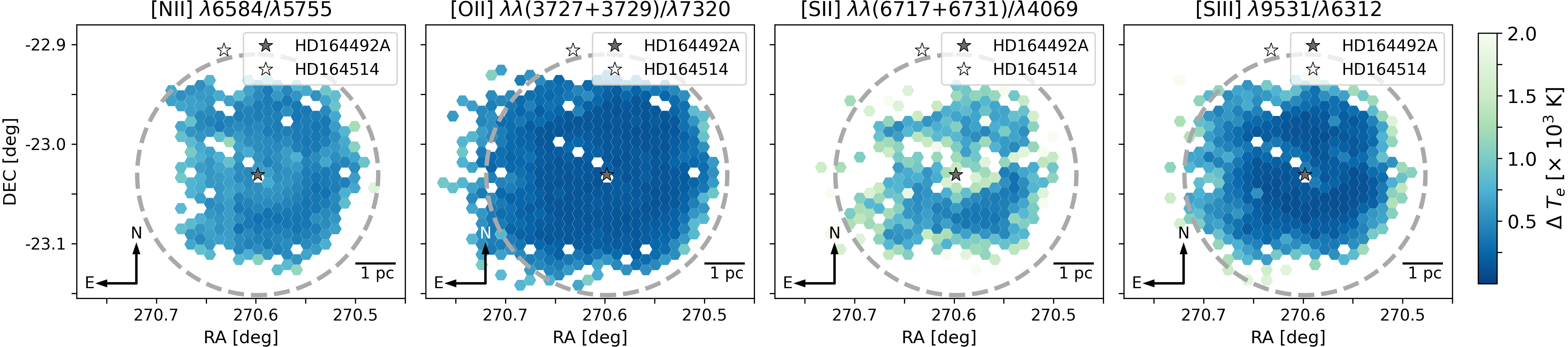}
    \caption{Uncertainties of the electron temperature maps shown in Fig.~\ref{fig:temperature}. The position of HD\,164492A is marked with a grey star. 
    The extent of the measured Strömgren sphere (Sect.~\ref{sec:Q0}) is shown by a grey dashed circle.}
    \label{fig:temperature_error}
\end{figure*}

\begin{figure*}
    \centering
    \hspace{-0.32cm}
    \includegraphics[width=0.66\linewidth]{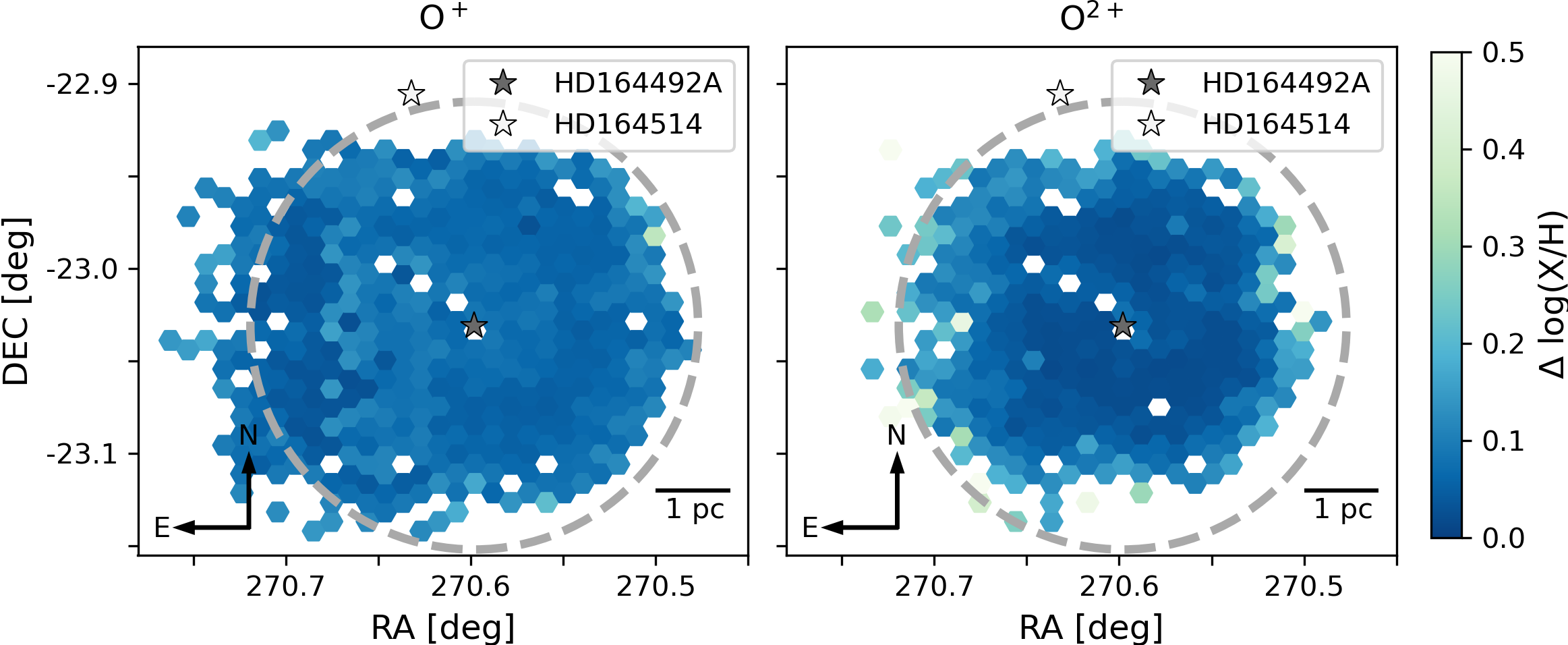}
    \caption{Uncertainties of the oxygen abundance maps shown in Fig.~\ref{fig:Oabundances}. The position of HD\,164492A is marked with a grey star.
    The extent of the measured Strömgren sphere (Sect.~\ref{sec:Q0}) is shown by a grey dashed circle.}
    \label{fig:Oabundance_error}
\end{figure*}

In Figs.~\ref{fig:reddening_error} to \ref{fig:Oabundance_error}, we present the uncertainty maps for dust extinction (Fig.~\ref{fig:reddening}), electron densities (Fig.~\ref{fig:density}), electron temperatures (Fig.~\ref{fig:temperature}) and oxygen abundances (Fig.~\ref{fig:Oabundances}).
The uncertainties for electron densities, electron temperatures, and oxygen abundances were derived from Monte Carlo Simulations as explained in detail in Sect.~\ref{sec:analysis}.
While the uncertainty of the dust extinction was derived as the standard deviation of all calculated correction factors derived from the various hydrogen line ratios (see Sect.~\ref{subsec:reddening}).

\twocolumn

\end{appendix}
\end{document}